\documentclass[twocolumn,tighten,numberedappendix]{aastex63}
\usepackage{amsmath}

\usepackage{lineno}

\def\simlt{\lower.5ex\hbox{$\; \buildrel < \over \sim \;$}}
\def\simgt{\lower.5ex\hbox{$\; \buildrel > \over \sim \;$}}

\def\beq{\begin{equation}}
\def\eeq{\end{equation}}
\def\ba{\begin{eqnarray}}
\def\ea{\end{eqnarray}}
\def\bB{\boldsymbol{B}}
\def\bE{\boldsymbol{E}}

\def\bj{\boldsymbol{j}}

\def\bv{\boldsymbol{v}}

\def\bk{\boldsymbol{k}}

\def\Sect{{\rm Section}}

\def\Eq{Equation}
\def\Eqs{Equations}

\def\sT{\sigma_{\rm T}}

\def\Bw{B_{\rm w}}

\def\rL{r_{\rm L}}

\def\E{{\cal E}}
\def\N{{\cal N}}

 \def\Grel{\Gamma_{\rm u|d}}

\def\RLC{R_{\rm LC}}

\def\bBw{\boldsymbol{B}_{\rm w}}

\def\bsh{\beta_{\rm sh}}

\def\be{\boldsymbol{e}}

\def\bsh{\beta_{\rm sh}}

\def\ep{\epsilon}
\def\omB{\omega_B}

\def\gsh{\gamma_{\rm sh}}

\def\Bbg{B_{\rm bg}}
\def\bBbg{\boldsymbol{B}_{\rm bg}}

\def\bEbg{\boldsymbol{E}_{\rm bg}}

\def\uu{\uparrow\uparrow}

\def\tomB{\tilde{\omega}_B}
\def\tB{\tilde{B}}

\def\vph{v_{\rm ph}}

\def\bb{\boldsymbol{\beta}}

\def\gD{\gamma_{\rm D}}
\def\bD{\beta_{\rm D}}

\def\bs{\beta_{\rm s}}
\def\gs{\gamma_{\rm s}}

\def\trho{\tilde{\rho}}

\def\xiin{\xi_{\rm i}}

\def\xiu{\xi^{\rm u}}

\def\xiinu{\xiin^{\rm u}}

\def\xiinc{\xiin^{\rm c}}

\def\trhou{\trho_{\rm u}}
\def\trhod{\trho_{\rm d}}

\def\c{\kappa}
\def\cu{\c_{\rm u}}
\def\cd{\c_{\rm d}}

\def\sigbg{\sigma_{\rm bg}}
\def\rhobg{\rho_{\rm bg}}
\def\nbg{n_{\rm bg}}

\def\xish{\xi_{\rm sh}}

\def\Abg{A_{\rm bg}}

\def\Em{E_0}
\def\Km{K_0}
\def\Rm{R_{\times}}
\def\rsh{r_{\times}}
\def\thsh{\theta_{\times}}
\def\Bm{B_{\times}}
\def\sigm{\sigma_{\times}}
\def\omegam{\omega_{\times}}

\def\Eu{E_{\rm u}}
\def\Bu{B_{\rm u}}
\def\gu{\gamma_{\rm u}}
\def\gd{\gamma_{\rm d}}
\def\betau{\beta_{\rm u}}
\def\betad{\beta_{\rm d}}
\def\Cu{C_{\rm u}}
\def\Cd{C_{\rm d}}

\def\uu{u_{\rm u}}

\def\tcross{t_{\rm cross}}
\def\cc{\c_{\rm c}}

\def\psic{\psi_{\rm c}}
\def\tc{t_{\rm c}}

\def\sigK{\sigma_\star}
\def\rK{r_\star}

\def\D{D}
\def\K{K}

\def\cv{\c_{\rm v}}
\def\tv{t_{\rm v}}

\def\gus{\gamma'_{\rm u}}
\def\gds{\gamma'_{\rm d}}

\def\sigu{\sigma_{\rm u}}

\def\Elost{\E_{\rm lost}}

\def\lm{\ell_\times}
\def\RF{R_{\rm F}}

\def\tt{s}
\def\R{{\cal R}}
\def\eps{\epsilon}

\def\Ed{E_{\rm d}}
\def\FF{{\cal F}}
\def\ww{w}

\def\omc{\omega_{\rm c}}
\def\epc{\ep_{\rm c}}
\def\lph{l_{\rm ph}}

\def\me{m}
\def\bA{\boldsymbol{A}}
\def\bAbg{\boldsymbol{A}_{\rm bg}}
\def\bAw{\bA_{\rm w}}
\def\Aw{A_{\rm w}}

\def\KF{{\cal K}}

\def\bu{\boldsymbol{u}}

\def\rvac{r_{\rm vac}}

\def\drK{\dot{r}_\star}

\def\Ep{E_{\rm p}}

\def\bq{\boldsymbol{n}}

\def\br{\boldsymbol{r}}

\def\eph{\varepsilon_\star}
\def\esc{\varepsilon_{\rm sc}}

\def\sigsc{\sigma_{\rm sc}}
\def\tausc{\tau_{\rm sc}}
\def\Lsc{L_{\rm sc}}

\def\tBu{\tilde{B}_{\rm u}}

\def\dEe{\dot{\E}_e}

\def\gMHD{\gamma_{\rm MHD}}

\def\xitrough{\xi_0}
\def\xidec{\xi_{\rm dec}}

\newbox\grsign \setbox\grsign=\hbox{$>$} \newdimen\grdimen \grdimen=\ht\grsign
\newbox\simlessbox \newbox\simgreatbox \newbox\simpropbox
\setbox\simgreatbox=\hbox{\raise.5ex\hbox{$>$}\llap
     {\lower.5ex\hbox{$\sim$}}}\ht1=\grdimen\dp1=0pt
\setbox\simlessbox=\hbox{\raise.5ex\hbox{$<$}\llap
     {\lower.5ex\hbox{$\sim$}}}\ht2=\grdimen\dp2=0pt
\setbox\simpropbox=\hbox{\raise.5ex\hbox{$\propto$}\llap
     {\lower.5ex\hbox{$\sim$}}}\ht2=\grdimen\dp2=0pt
\def\simgt{\mathrel{\copy\simgreatbox}}
\def\simlt{\mathrel{\copy\simlessbox}}

\begin{document}

\title{Monster radiative shocks in the perturbed magnetospheres of neutron stars}

 \email{amb@phys.columbia.edu}

\author[0000-0001-5660-3175]{Andrei M. Beloborodov}
\affiliation{Physics Department and Columbia Astrophysics Laboratory, Columbia University, 538  West 120th Street New York, NY 10027,USA}
\affil{Max Planck Institute for Astrophysics, Karl-Schwarzschild-Str. 1, D-85741, Garching, Germany}

\begin{abstract}
Magnetospheres of neutron stars can be perturbed by star quakes, interaction in a  binary system, or sudden collapse of the star. The perturbations are typically in the kHz band and excite magnetohydrodynamic waves. We show that compressive magnetospheric waves steepen into monster shocks, possibly the strongest shocks in the universe. The shocks are radiative, i.e. the plasma energy is radiated before it crosses the shock. As the kHz wave with the radiative shock expands through the magnetosphere, it produces a bright X-ray burst. Then, it launches an approximately adiabatic blast wave, which will expand far from the neutron star. These results suggest a new mechanism for X-ray bursts from magnetars and support the  connection of magnetar X-ray activity with fast radio bursts. Similar shocks may persistently occur in magnetized neutron-star binaries, generating an X-ray precursor of the binary merger. Powerful radiative shocks are also predicted in the magnetosphere of a neutron star when it collapses into a black hole, producing a bright X-ray transient.
\end{abstract}

 \keywords{
X-ray transient sources (1852);
Neutron stars (1108);
Magnetars (992);
Radiative processes (2055);
Radio bursts (1339);
Plasma astrophysics (1261)
}


\section{Introduction}

Magnetospheres of neutron stars are formed by plasma immersed in a strong magnetic field $\Bbg$.  They have a high magnetization parameter,
\beq
    \sigbg\equiv \frac{\Bbg^2}{4\pi \rhobg c^2}\gg 1,
\eeq
where $\rhobg$ is the plasma mass density and $c$ is the speed of light; subscript ``bg'' stands for ``background'' for waves investigated in this paper. The closed magnetosphere is approximately dipole at radii $r$ much greater than the star radius $R_\star$. It ends and a magnetized wind begins near the light-cylinder $\RLC=c/\Omega$, where  $\Omega$ is the star rotation rate. This basic picture is confirmed by extensive studies of pulsars \citep{Philippov22}.

\subsection{Perturbations of neutron star magnetospheres}

In some cases, the magnetosphere becomes significantly perturbed. In particular, quakes in magnetars launch magnetospheric waves \citep{Blaes89,Thompson96,Bransgrove20}. Some mechanism quickly dissipates the waves, generating X-ray bursts that are observed as the main form of magnetar activity \citep{Kaspi17}. Significant perturbations are also expected in tight neutron-star binaries, where the two magnetospheres interact with each other. A huge disturbance of a neutron star magnetosphere occurs when the star collapses into a black hole (\citealt{Lehner12}). In all these cases, the excited waves are typically in the kHz band. 

\begin{figure*}[t]
\begin{center}
\includegraphics[width=0.85\textwidth]{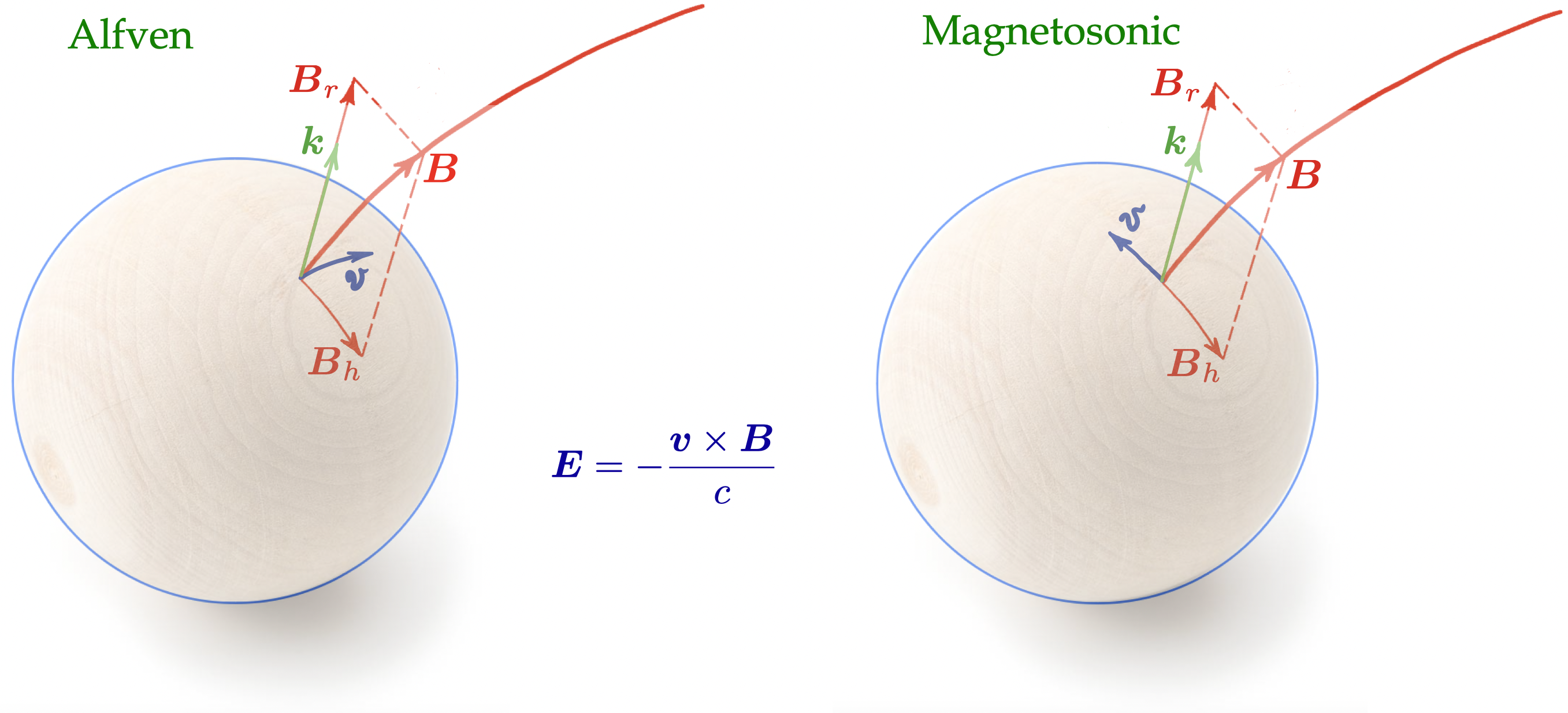} 
\vspace*{-4mm}
\caption{Excitation of magnetospheric waves by a star quake with $\omega\gg c/R_\star$ and a vertical (radial) wavevector $\bk$. The wave polarization is set by the direction of the electric field $\bE=\bB\times\bv/c$ in the crustal shear oscillation with a horizontal velocity $\bv$. Left: crustal motion with $\bv\parallel\bq\equiv \bk\times\bBbg$ excites an Alfv\'en wave $\bE\perp\bq$. Right: $\bv\perp\bq$ excites a magnetosonic wave $\bE\parallel\bq$. The wave amplitude is small, so $\bB\approx\bBbg$ near the star.}
\end{center}
\label{fig:quake}
 \end{figure*}

Such perturbations are well described by magnetohydrodynamics (MHD), which supports waves of the perturbed magnetic field and electric field $\bE\perp\bBbg$. These MHD modes can have wavevectors $\bk$ in any direction and can be of two types: (1) shear Alfv\'en waves with $\bk\cdot\bE\neq 0$, and (2) compressive waves with $\bE\perp\bk$, so-called ``fast magnetosonic modes.'' Both modes propagate with an ultrarelativistic group speed,
\beq
   \frac{v_{\rm gr}}{c}\approx 1-\sigbg^{-1}. 
\eeq
Phase speed $\vph=\omega/k$ of the magnetosonic waves is also close to $c$ while Alfv\'en waves have $v_{\rm ph}\approx c\cos\alpha$, where $\alpha$ is the angle between $\bk$ and $\bBbg$. A detailed discussion of waves in $e^\pm$ plasma in a strong background magnetic field $\Bbg$, including two-fluid and single-fluid (MHD) descriptions, is found in \cite{Arons86}.

\subsection{MHD waves from quakes}
\label{quake}

Substantial work was devoted to the relativistic Alfv\'en waves excited by magnetar quakes (e.g. \citealt{Thompson98,Troischt04,Yuan20,Nattila22}), and little attention was given to the magnetosonic waves. In fact, quakes can excite both modes. 

Neutron star quakes involve shear oscillations of the crust with horizontal displacements $\delta\br$. Shear waves inside the neutron star crust propagate with speed $v_{\rm sh}=\omega/k_{\rm sh}\approx 10^8$\,cm/s \citep{Blaes89}. Their characteristic lowest frequency is $\omega\sim v_{\rm sh}/h\sim 10^4\,$rad/s, where $h$ is the hydrostatic scale height. A large fraction of the quake energy may be at much higher frequencies, e.g. $\omega\sim 10^5$\,rad/s. The crustal waves leak to the magnetosphere with a transmission coefficient ${\cal T}\sim 0.1\, \omega_5^{3/5} B_{15}^{2/5}$ \citep{Blaes89,Bransgrove20}. 

To quickly see that quakes can emit both Alfv\'en and magnetosonic  modes, one can consider crustal oscillations with $\omega\gg c/R_\star$ and a wavevector that is exactly radial in a region of the stellar surface (Figure~1). Such oscillations satisfy $\partial_\phi=\partial_\theta=0$ in spherical coordinates $r,\theta,\phi$, and this symmetry is preserved during wave transmission, so the emitted magnetospheric wave also has a radial $\bk$. This simple setup is reduced to a locally plane-parallel transmission problem; it was used by \cite{Blaes89} to study how quakes with $\delta\br\parallel\bk\times\bBbg$ excite Alfv\'en waves in the magnetosphere. It is easy to see that quakes with $\delta\br\nparallel\bk\times\bBbg$ emit magnetosonic modes. Polarization of the emitted wave is set by the electric field in the crustal oscillation, $\bE=\bBbg\times\bv/c$ (where $\bv=-i\omega\delta\br$), and its type is determined by the orientation of $\bE$ relative to the vector $\bq\equiv \bk\times\bBbg$. It is a pure Alfv\'en mode if $\bE\perp \bq$ and a pure magnetosonic wave if $\bE\parallel \bq$. The condition for magnetosonic excitation $\bE\cdot\bq\neq 0$ may be stated as 
\beq
   (\bBbg\times\bv)\cdot (\bk\times \bBbg)  =  (\bBbg\cdot\bv)(\bBbg\cdot\bk) \neq 0.
\eeq
The transmitted quake power is partitioned between the Alfv\'en and magnetosonic modes as $E_{\rm A}^2/E^2_{\rm ms}=\tan^2\!\psi$ where $\psi$ is the angle between the quake motion $\bv$ and the horizontal component of the magnetic field $\bBbg^{\rm h}$.

A complete picture of wave emission is complicated by several additional effects:
\\
(1) Besides the vertical (radial) wavevector $k_r$, crustal waves have a horizontal component $k_{\rm h}$ \citep{Bransgrove20} and excite non-radial magnetospheric waves. As an example, \cite{Yuan20} examined wave generation in a uniform oblique $\bBbg$ attached to a horizontal conducting surface perturbed by axisymmetric horizontal shear. They showed that the shear excites a mixture of Alfv\'en and magnetosonic waves. These calculations should be extended in the future to include a realistic density profile of the crust.
\\
(2) Waves of lower frequencies $\omega\simlt c/R_\star$ develop at larger radii $R_{\rm w}\sim R_\star+c/\omega$. In the region $R_\star<r<R_{\rm w}$ the magnetosphere adjusts to the surface oscillation in a quasi-static manner and, effectively, the magnetospheric footpoints in the wave are relocated from the stellar surface to the sphere of radius $r\approx R_{\rm w}$. 
The quasi-static deformation of the magnetosphere at $r\simlt R_{\rm w}$ differs from the crustal deformation and needs further investigation of its compressive component that could drive a small-amplitude magnetosonic wave. For example, a strongly twisted magnetosphere in axisymmetric equilibrium responds to additional surface shear $\partial_\theta v_\phi\neq 0$ by inflating at $r\simlt R_{\rm w}$ \citep{Parfrey13}. 
\\
(3) The twisted magnetospheres near magnetars have significant spatial variations of the toroidal magnetic field component $B_\phi$, so vector $\bq=\bk\times\bBbg$ can change its direction on a scale comparable to $r$. Then, any attempt to launch a pure Alfv\'en mode at $r\sim R_{\rm w}\sim c/\omega$ (for $\omega\simlt c/R_\star$) will inevitably generate a mixture of the Alfv\'en and magnetosonic waves because the two linear modes have different refraction indices $c/v_{\rm ph}$ and will be unable to adiabatically track the changing local $\bq$ until propagating to $r\gg c/\omega$.

Pure Alfv\'enic excitations occur in the simple case of axisymmetric quakes with azimuthal $\bv$ and an untwisted background magnetosphere (then $k_\phi=0$ and $\Bbg^\phi=0$,  so $\bq=\bk\times\bBbg\parallel\bv$). Even in this case, the emitted Alfv\'en waves can convert to magnetosonic waves because of nonlinear effects. Alfv\'en waves have $\bv_{\rm gr}\parallel \bBbg$, so they are ducted along the closed magnetic loops, and the nonlinear conversion peaks when the wave reaches the top of the loop \citep{Yuan21}. In addition, at later times, the bouncing Alfv\'en waves trapped in the loop develop a nonlinear turbulent cascade which emits magnetosonic waves (e.g. \citealt{Li19}). 

Besides sudden quakes, the crust may flow plastically and slowly twist a magnetospheric flux bundle to an instability threshold. Three-dimensional simulations of this process demonstrate that relaxation of the unstable flux bundle generates fast magnetosonic waves in the magnetosphere \citep{Mahlmann23}.

\subsection{Shocks}

In the limit of high $\sigbg$, the emitted magnetosonic waves are equivalent to electromagnetic radio waves propagating without coupling to the magnetized plasma: the oscillating $\bE\perp\bBbg$ drives a tiny electric current, negligible compared to the displacement current $\partial_t\bE/4\pi$. Therefore, compressive perturbations of the magnetosphere are usually assumed to propagate as vacuum electromagnetic waves superimposed linearly with the background field $\bBbg$, and freely escape.\footnote{It was also proposed that QED effects in ultrastrong $\Bbg$ could steepen a high-frequency wave into a shock \citep{Heyl05}. Such effects are absent for kHz waves.}
The linear propagation would imply no shock formation.

However, this simple picture is safe only for low-amplitude waves, $|\bE|\ll\Bbg$, i.e. near the neutron star where $\Bbg$ is strong. As the wave expands to larger radii $r$, the dipole background field decreases as $\Bbg\propto r^{-3}$ while the wave amplitude $\Em$ decreases as $r^{-1}$. Their ratio $\Em/\Bbg$ grows as $r^2$ and eventually approaches unity. It is easy to see that the linear propagation of an oscillating wave  becomes impossible when 
\beq
\label{eq:break}
   2\Em\sin\alpha>\Bbg, 
\eeq
where $\alpha$ is the angle between $\bBbg$ and the wavevector $\bk$. Indeed, the linear superposition of $\bBbg$ with the vacuum electromagnetic wave ($\bE_{\rm w}=\bE$ and $\bBw=\bB-\bBbg$) gives
\beq
\label{eq:B_E}
    B^2-E^2 = \Bbg^2+2\bBbg\cdot\bBw,
\eeq
where we used $\bE_{\rm w}^2=\bBw^2$. MHD description breaks if $B^2<E^2$. This occurs when the condition~(\ref{eq:break}) is met.

\begin{figure}[t]
\begin{center}
\includegraphics[width=0.4\textwidth]{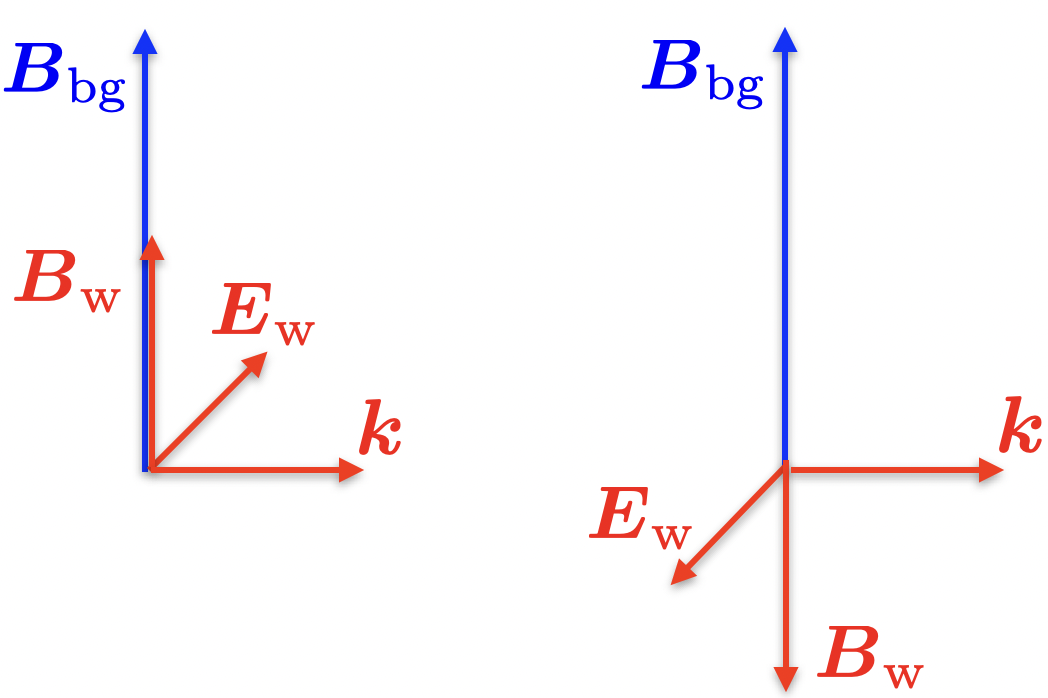} 
\caption{
A magnetosonic wave with wavevector $\bk\perp\bBbg$ at two oscillation phases: when the wave field $\bBw=\bB-\bBbg$ is aligned with $\bBbg$ (left) and anti-aligned with $\bBbg$ (right). The wave electric field $\boldsymbol{E}_{\rm w}=\bE$ is always orthogonal to $\bBbg$. The quantity $B^2-E^2$ reaches zero when $\bBw=-\bBbg/2$. At this moment, the plasma drift speed $\bv=c\,\bE\times\bB/B^2$ approaches $c$, so it experiences ultra-relativistic acceleration in the direction opposite to the wave propagation direction.}
\end{center}
\label{fig:wave}
 \end{figure}

For instance, consider the equatorial plane of a dipole magnetosphere. Here, a spherical wave front propagates with $\bk\perp \bBbg$ ($\alpha=\pi/2$) and the magnetic field perturbation $\bBw$ oscillates along $\bBbg$ (Figure~2). The linear superposition wave+background gives minimum $B^2-E^2$ at $B_{\min}=\Bbg-\Em$, reaching $B^2-E^2=0$  when $\Em=\Bbg/2$. Note that the plasma oscillates  with the drift speed $\bv/c=\bE\times\bB/B^2$, and $E^2\rightarrow B^2$ corresponds to $|\bv|\rightarrow c$. This implies a runaway growth of the plasma kinetic energy, and the MHD wave can no longer be approximated as a vacuum electromagnetic wave. Such energy conversion is a general effect of $E^2\rightarrow B^2$ (it also happens in Alfv\'en waves \citep{Li19,Li21}; see \cite{Levinson22} for a recent discussion). In magnetosonic waves, energy conversion prevents reaching $E^2=B^2$ by steepening the wave into a shock. The appearance of shocks at $\Em=\Bbg/2$ was previously noted in the context of  waves in pulsar winds \citep{Lyubarsky03}. A fully kinetic (local-box) particle-in-cell simulation of a magnetosonic wave propagating through a decreasing $\Bbg$ has recently been performed by \cite{Chen22}. Their simulation demonstrates the sudden steepening of the wave into a shock when $E^2$ approaches $B^2$.

Waves consisting of half oscillation with $\bBw\cdot\bBbg>0$ and no part with $\bBw\cdot\bBbg<0$ would never face the $E^2=B^2$ limit. One may think that in this case the wave avoids shock formation. However, as explained below, such half-waves  also steepen into shocks, although this occurs gradually, at a larger radius. This gradual steepening creates a forward shock, launching an accelerating blast wave that expands far beyond the magnetosphere.

 \subsection{This paper}

Since shocks appear to be a generic outcome of magnetosonic perturbations, a few questions arise: How strong are the shocks? What fraction of the original wave energy gets dissipated by the shock? What fraction of the dissipated energy is radiated? 

These questions can be answered by solving a well-defined MHD problem with a simple initial condition: launch a spherically expanding magnetosonic wave with an initial amplitude $E_0/\Bbg\ll 1$ so that initially (at small radii) it behaves as a vacuum electromagnetic wave. Then, track its expansion through the dipole magnetosphere and examine how it steepens into a shock and propagates afterward. This problem is solved in the present paper. We will show that the plasma Lorentz factor $\gamma$ in the magnetospheric shocks caused by $E^2\rightarrow B^2$ reaches huge values $\gamma\propto\sigbg$, likely making them the strongest shocks in the universe. Therefore, we call them ``monster shocks.'' They differ from normal collisionless plasma shocks because they are highly radiative: we will show that the plasma approaching the monster shock radiates its kinetic energy before forming the downstream flow.

Tracking the evolution of magnetospheric waves with shock formation presents an interesting technical challenge. The magnetization parameter $\sigbg$ at radii of main interest can exceed $10^{10}$ (\Sect~\ref{bg}). The large $\sigbg$ is usually replaced by $\sigbg\rightarrow\infty$, which corresponds to taking the force-free electrodynamics (FFE) limit of MHD. However, FFE cannot describe shock formation. FFE neglects plasma inertia, so its wave modes propagate with exactly speed of light and cannot steepen.

Since FFE does not provide a suitable framework, one has to examine the problem  using the full MHD equations. The equations state conservation of energy and momentum and could be solved numerically with customary discretization methods. However, in practice such methods fail at high magnetization $\sigma$ (most existing MHD codes have to keep $\sigma<100$ to avoid numerical issues). We take a different approach: we solve the MHD equations along characteristics. It provides an efficient method for both finding and understanding the solution at arbitrarily high $\sigbg$. The solution is easiest to find for short waves, with wavelength $\lambda\ll r$.

Calculations in this paper are performed for axisymmetric wave packets ($\partial_\phi=0$). Setting $\partial_\phi=0$ should also be a good approximation more generally for waves far from their source, where wavevectors $\bk$ are radial. The wave evolution along the radial ray is controlled by the local $\bBbg$ encountered along the ray; we consider the simple case of a dipole $\bBbg$.

Formulation of the problem and our method for solving it are described in \Sect~\ref{NW}. \Sect~\ref{shock} explains how we track shocks in MHD waves and then \Sect~\ref{numerical} presents the full numerical simulation, performed for an equatorial wave in a dipole magnetosphere. Remarkably, the problem also admits a complete analytical description, which is given in \Sect~\ref{analytical}. Emission from the monster shock is discussed in \Sect~\ref{emission}. The results are discussed in \Sect~\ref{discussion}.


\section{Nonlinear magnetosonic waves}
\label{NW}

MHD fluid is described by the plasma mass density $\rho$, velocity $\bv=c\bb$, magnetic field $\bB$, and electric field $\bE$. We wish to investigate magnetosonic waves launched in a dipole magnetosphere and find their nonlinear evolution. Term ``nonlinear'' here means that (a) the wave oscillation is not necessarily small compared to the background field $\bBbg$, and (b) the wave can be strongly deformed during its evolution and form shocks.

\subsection{Unperturbed background magnetosphere}
\label{bg}

The wave will be shown to experience strong evolution at radii $r$ much greater than the neutron star radius $R_\star\sim 10^6$\,cm, but well inside the light cylinder $\RLC$ (all known magnetars in our galaxy have $\RLC\simgt 10^{10}\,$cm). Our calculations below use the background at radii $R_\star\ll r\ll\RLC$, the magnetosphere is approximated as dipole, and its rotation is neglected. Thus, for simplicity, we neglect any twists of the outer magnetosphere, i.e. assume $\Bbg^\phi\approx 0$ at radii of interest. The electromagnetic field of a static dipole magnetosphere is
\beq
  \bEbg=0, \qquad \bBbg=\frac{2\mu\cos\theta}{r^3}\be_{r}+\frac{\mu\sin\theta}{r^3}\be_{\theta},
\eeq
where $(\be_r,\be_{\theta},\be_{\phi})$ is the normalized basis in spherical coordinates $r,\theta,\phi$ with the polar axis along the magnetic dipole moment $\boldsymbol{\mu}$. Magnetars have $\mu\sim 10^{33}\,$G~cm$^3$.

Plasma in the unperturbed static magnetosphere will be approximated by 
\beq 
  \rhobg = \frac{\N\me}{r^3}, \qquad  \bv_{\rm bg}\approx 0,
\eeq
where the dimensionless parameter $\N\equiv \nbg r^3$ is approximately constant with $r$ \citep{Beloborodov20}; $\me$ is the particle ($e^\pm$) mass, and $\nbg=\rhobg/\me$. The lowest $\N\sim 10^{31}\,\mu_{33}\,(\Omega/{\rm rad\,s}^{-1})$ is set by the Goldreich-Julian density. The more typical $\N$ for magnetars is much higher, $\N\sim 10^{37}$, due to strong electric currents near the star accompanied by $e^\pm$ pair creation with a high multiplicity \citep{Beloborodov21b}.\footnote{The created pairs flow along the magnetic field lines with a speed controlled by the magnetar's radiation field  \citep{Beloborodov13a}. In the outer closed magnetosphere, the plasma accumulates and annihilates at $\theta\approx\pi/2$, in a layer of high $e^\pm$ density controlled by annihilation balance and the thickness of the annihilation layer. The $e^\pm$ outflow terminated by the annihilation sink at $\theta=\pi/2$ may have $\N\sim 10^{37}$ at all $\theta$ except the thin equatorial layer.}
The magnetization parameter of the background plasma $\sigbg=\Bbg^2/4\pi\rhobg c^2$ is
\beq
\label{eq:sigbg}
   \sigbg \approx \frac{\D}{r^3} \sim 10^{10}\mu_{33}^2\N_{37}^{-1} r_8^{-3},
   \quad\; D\equiv \frac{\mu^2}{4\pi \N \me c^2}.
\eeq

The magnetosphere at $r>10^7\,$cm is populated with mildly relativistic $e^\pm$, as they are decelerated by drag exerted by the magnetar radiation \citep{Beloborodov13a}. We neglect these background motions; thus,
the plasma energy density (including rest mass) in the unperturbed background ahead of the wave is approximated as $\rhobg c^2$.

\subsection{Axisymmetric waves}

We will examine the radial expansion of an axisymmetric magnetosonic wave. Its electric field $\bE$ is perpendicular to both $\bBbg$ and the (radial) propagation direction, so vector $\bE$ oscillates along $\phi$. It is convenient to define scalar $E$ by
\beq
   \bE=-E\be_{\phi}, \qquad E\equiv -E_{\phi}.
\eeq
The wave magnetic field is perpendicular to $\bE$ and oscillates in the poloidal plane,
\beq 
   \bB=B_r\be_r + B_{\theta} \be_{\theta}.
\eeq 
Any axisymmetric wave can be described by a toroidal electromagnetic potential,
\beq
  \bA=A(t,r,\theta)\,\be_{\phi},
  \quad\;  c\bE=-\partial_t \bA, \quad\; \bB=\nabla\times \boldsymbol{A}.
\eeq
This class of fields includes the static dipole background as a special case, with  
\beq
   \bAbg(r,\theta)=\frac{\mu\sin\theta}{r^2}\,\be_{\phi}.
\eeq

Charge density remains zero in magnetosonic waves, $4\pi \rho_e = \nabla\cdot \bE = 0$. Electric current density $\bj$ satisfies the Maxwell equation $4\pi\bj=-\partial_t\bE + c\,\nabla\times\bB$. In axisymmetric waves the current is toroidal:
\beq
\label{eq:j}
  4\pi\bj = \left\{\partial_t E +\frac{c}{r}\left[\partial_r(rB_\theta)-\partial_\theta B_r\right]\right\} \be_\phi.
\eeq
Note also that $\bE$ are $\bB$ are related by induction equation $\partial_t \bB=-c\nabla\times \bE$ (the identity $\partial_t\nabla\times\bA=\nabla\times\partial_t\bA$):
\beq
\label{eq:induction}
  \partial_t(rB_\theta)=-c\,\partial_r(rE), \quad \partial_t(rB_r)=\frac{c\,\partial_\theta(E\sin\theta)}{\sin\theta}.
\eeq

As a concrete example, we will consider a wave launched with an initial sine profile
\beq 
\label{eq:initial}
   E(\xi)=\Em\sin(\omega\xi), \qquad \xi=t-\frac{r}{c}.
\eeq
One full oscillation corresponds to $0<\xi<\lambda/c=2\pi/\omega$.

\subsection{Short waves ($\lambda\ll r$)}
\label{sw}

The condition $\lambda\ll r$ considerably simplifies the wave propagation problem. In particular, it leads to a simple expression for current $\bj$, which will be used in \Sect~\ref{NWE} to describe the wave-plasma interaction. 

Let us define the wave potential $\bAw\equiv (A-\Abg)\be_\phi$. It determines the wave fields and $\bj$:
\beq
  \bBw\equiv \bB-\bBbg=\nabla\times\bA_{\rm w}, \qquad c\bE=-\partial_t \bA_{\rm w},
\eeq
\beq
\label{eq:j1}
  \bj = \left[ 
 \partial_t(rE)+c\,\partial_r(r\Bw^\theta)-c\,\partial_\theta \Bw^r
  \right] \frac{\be_\phi}{4\pi r}.
\eeq
For short waves, it is helpful to view fields as functions of $t,\xi,\theta$ instead of $t,r,\theta$. The fast oscillation then becomes isolated in the coordinate $\xi$. Differential equations can be rewritten in variables $t,\xi,\theta$ using
\beq
   \left.\partial_t \right|_r = \left.\partial_t  \right|_\xi + \left.\partial_\xi  \right|_t, 
   \qquad \left.\partial_r  \right|_t = -\left.\partial_\xi  \right|_t.
\eeq
Using the $\theta$ component of induction equation~(\ref{eq:induction}),
\beq
\label{eq:ind1}
    \partial_\xi(rE)_t = \partial_t(r\Bw^\theta)_\xi +\partial_\xi (r\Bw^\theta)_t,
\eeq
we rewrite \Eq~(\ref{eq:j1}) as 
\beq
\label{eq:j2}
  \bj = \left[ \partial_t(rE+r\Bw^\theta)_\xi-c\,\partial_\theta \Bw^r \right] \frac{\be_\phi}{4\pi r}.
\eeq
In short waves, the oscillation of $\Aw$ with $\xi$ is much faster than its variation with $t$ or $\theta$ at fixed $\xi$. Hence,
\beq
   B_{\rm w}^\theta = \frac{1}{c\,r}\,\partial_\xi(r\Aw)\gg B_{\rm w}^r,
\eeq
as $\Bw^r$ does not contain the large derivative $\partial_\xi(r\Aw)$. Note also that $rE = \partial_\xi(r\Aw)_t + \partial_t(r\Aw)_\xi$, and 
\beq
   rE-r\Bw^\theta=\frac{1}{c}\,\partial_t(r\Aw)_\xi\ll rE.
\eeq 
Therefore, the expression for $\bj$ simplifies to
\beq
\label{eq:j_sw}
   \bj = \frac{\partial_t(rE)_\xi}{2\pi r}\,\be_\phi.
\eeq

There is also a useful relation between plasma density $n=\rho/m$ and velocity $\bv$ in short waves. In general, $n$ and $\bv$ satisfy the continuity equation $\partial_t n +\nabla\cdot(n\bv)=0$. In the short-wave limit, it simplifies to 
\beq
\label{eq:sw}
   F_\xi\equiv (c-v_r)n=const=c \nbg. 
\eeq
$F_\xi$ is the particle flux through the surface of $\xi=const$. It is uniform across the wave packet, and evolves as the packet propagates to larger radii: $F_\xi=c\nbg\propto r^{-3}$.

\subsection{Wave-plasma interaction}
\label{NWE}

We wish to find the evolution of the wave profile $E(\xi)$. It evolves because the electromagnetic field exchanges energy and momentum with the plasma oscillating in the wave. The plasma motion is described by four-velocity,
\beq
   u^\mu=(\gamma,\bu), \quad\; \bu\equiv\gamma\bb, \quad\; \gamma^2=1+u^2=
   \frac{1}{1-\beta^2}, 
\eeq
where $\bb=\bv/c$. The equation of motion is 
\beq
\label{eq:momentum}
   \rho c^2\,\frac{d\bu}{dt}=\bj\times\bB,
\eeq
with the derivative $d/dt$ taken along the fluid streamline: $d/dt=\partial_t+\bv\cdot\nabla$. Taking the scalar product of both sides with $\bb$, and using $\bE+\bb\times\bB=0$, one obtains 
\beq
\label{eq:energy}
   \rho c^2\, \frac{d\gamma}{dt} =  \bE\cdot\bj.
\eeq
This equation expresses energy conservation. Note that using $\rho c^2=n\me c^2$ in \Eq~(\ref{eq:momentum}) assumes a negligible contribution of internal (thermal) energy to the plasma inertial mass.  As explained below, this is a reasonable approximation, because significant wave-plasma interaction happens where the plasma has a low temperature and a high $\gamma$.

Substitution of \Eq~(\ref{eq:j_sw}) into \Eq~(\ref{eq:energy}) gives
\beq
\label{eq:energy_sw}
  -\frac{E}{2\pi r^2}\, \partial_t(rE)_\xi = \rho c^2\,\frac{d\gamma}{dt}
  = \rhobg c^2 \,\frac{d\gamma}{d\xi},
\eeq
where we used $d\xi=dt-dr/c=(1-\beta_r)dt$ along the fluid streamline and $\rho(1-\beta_r)=\rhobg$ (\Eq~\ref{eq:sw}). The derivative $d\gamma/d\xi$ written out in coordinates $x^{\alpha}=(t,\xi,\theta,\phi)$ takes the form,
\beq
   \frac{d\gamma}{d\xi} = \frac{dx^{\alpha}}{d\xi}\,\partial_{\alpha} \gamma 
   = \frac{\partial_t\gamma}{1-\beta_r} + \partial_\xi\gamma+\frac{v_\theta\,\partial_\theta\gamma}{r(1-\beta_r)}.
\eeq 
One might expect the terms with $\partial_t\gamma$ and $\partial_\theta\gamma$ to be small compared to $\partial_\xi\gamma$ by the factor of $\sim\lambda/r\ll 1$. However, nonlinear evolution can make the term with $\partial_t\gamma$ important. As shown below, this happens when $\gamma^3\simgt\sigbg$. 

Now we can write the energy equation in its final form:
\beq
\label{eq:NWE}
    \partial_t(r^2E^2) =-4\pi r^2\rhobg c^2\left[\partial_\xi\gamma+\frac{r\partial_t\gamma+\beta_\theta\partial_\theta\gamma}{r(1-\beta_r)}\right],
\eeq 
where derivative $\partial_t$ is taken at fixed $\xi,\theta,\phi$, i.e. along the radial ray $r=ct+const$. This equation connects the evolution of $E(t,\xi,\theta)$ and $\gamma(t,\xi,\theta)$.

\Eq~(\ref{eq:NWE}) has a simple intuitive interpretation. Multiplication of both sides by an infinitesimal $\delta\xi$ gives 
\beq
\label{eq:en0}
   \left.\partial_t\delta\E\right|_\xi 
   = -\dot{\N}_\xi \me c^2 \delta\gamma.
\eeq
Here, $\dot{\N}_\xi\equiv 4\pi r^2 F_\xi$ is the isotropic equivalent of the particle flux $F_\xi$, and $\delta\gamma=(d\gamma/d\xi)\delta\xi$ is the change of $\gamma$ along the fluid streamline as it crosses $\delta\xi$; it determines the energy gain of the plasma crossing $\delta\xi$ per unit time, $\dot{\N}_\xi \me c^2 \delta\gamma$. The quantity $\delta\E=c\,r^2 E^2\,\delta\xi$ is the electromagnetic wave energy (isotropic equivalent) contained in the infinitesimal part $\delta\xi$ of the wave profile. This interpretation of $\delta\E$ is consistent with the energy density $U_{\rm w}=(E^2+B_{\rm w, \theta}^2+B^2_{{\rm w},r})/8\pi\approx E^2/4\pi$ for the electromagnetic wave in the short-wave limit. 

Thus, \Eq~(\ref{eq:NWE}) merely states that the wave profile $r^2E^2(\xi)$ evolves by exchanging energy with the plasma. The plasma Lorentz factor $\gamma(\xi)$ oscillates, and $E(\xi)$ becomes deformed because the plasma passing through the wave receives energy at some $\xi_1$ and returns it to the electromagnetic field at another $\xi_2$. $E(\xi)$ is systematically reduced where $d\gamma/d\xi>0$ and increased where $d\gamma/d\xi<0$, leading to the steepening of $E(\xi)$. 

The FFE limit of MHD would correspond to setting $\rhobg\rightarrow 0$, so that the r.h.s. of \Eq~(\ref{eq:NWE}) vanishes. Hence, the FFE solution is $rE=const$ at $\xi=const$. This wave excites no electric current $\bj$, as can be verified using \Eq~(\ref{eq:j}), and so the FFE limit gives a simple vacuum wave superimposed on $\bBbg$. 

\Eq~(\ref{eq:NWE}) is sufficient to find the evolution of MHD waves if they drive pure radial plasma motions, as happens in the equatorial waves described below. Then, energy conservation (or the radial component of momentum equation~(\ref{eq:momentum})) contains the complete dynamical information. Wave propagation outside the equatorial plane ($\theta\neq\pi/2$) involves additional $\theta$-motions governed by the $\theta$ component of \Eq~(\ref{eq:momentum}).

\subsection{Equatorial waves}
\label{equatorial}

Main features of shock development will be shown by tracking waves at $\theta=\pi/2$ that are symmetric about the equatorial plane. Symmetry implies $u^\theta=0$ and $B^r=0$ at $\theta=\pi/2$. The MHD fluid then oscillates with velocity
\beq
  \bb = \frac{\bE\times\bB}{\bB^2}=\beta\,\be_r.
\eeq
We will use the following notation:
\beq
    \beta\equiv\beta_r, \qquad B\equiv B_{\theta},  \qquad E\equiv -E_{\phi}.
\eeq
These definitions imply $E^2=\bE^2$, $B^2=\bB^2$ at $\theta=\pi/2$ while $E$, $B$, and $\beta$ may be positive or negative. Note that 
\beq 
   \beta = \frac{E}{B},   \qquad   \gamma^2=\frac{B^2}{B^2-E^2}.
\eeq 
\Eq~(\ref{eq:NWE}) for the equatorial wave becomes
\beq
\label{eq:en2}
    \partial_t (r^2E^2) = -4\pi r^2 \rhobg c^2
     \left(\partial_\xi\gamma+\frac{\partial_t \gamma}{1-\beta}\right).
\eeq
Recall that the partial derivatives $\partial_t$ and $\partial_\xi$ are defined in variables $(t,\xi)$ and can act on any function $E(t,\xi)$, $\gamma(t,\xi)$, or $r(t,\xi)=c(t-\xi)$ entering \Eq~(\ref{eq:en2}). 

The equatorial wave problem can be reduced to one unknown function of $t,\xi$. Indeed, any short wave packet propagating into an initially unperturbed plasma has a useful feature: fluid compression $n/\nbg$ at any point in the packet is related to the local speed $\beta$ (\Eq~\ref{eq:sw}). The magnetic field is frozen in the fluid and compressed by the same factor, so
\beq
\label{eq:continuity_short}
   \frac{n}{\nbg}=\frac{\rho}{\rhobg}=\frac{B}{\Bbg}=(1-\beta)^{-1}.
\eeq 
The electric field $E=\beta B$ can be expressed as 
\beq
\label{eq:E}
  E=\frac{\Bbg\beta}{1-\beta}.
\eeq
Thus, all MHD quantities in a short wave are functions of $\beta$, and \Eq~(\ref{eq:en2}) can be recast so that it contains only derivatives of $\gamma$. This requires expressing $\partial_t(rE)$ in terms of $\partial_t\gamma$. Differentiating \Eq~(\ref{eq:E}), we find\footnote{The short-wave approximation for $E(\beta)$ (\Eq~\ref{eq:E}) is equivalent to $E=\Bw$. It can be used when expressing $\left.\partial_tE\right|_\xi$ in terms of $\left.\partial_t\beta\right|_\xi$. Note, however, that setting $E=\Bw$ would not be safe in expressions containing the large derivative $\partial_\xi$. In particular, $\partial_\xi(\Bw-E)$ in short waves is of the same order as $\left.\partial_tE\right|_\xi$.}
\beq 
\label{eq:E_gamma}
  \partial_t(rE)=-2cE+\frac{r\Bbg\,\partial_t\gamma}{(1-\beta)^2\gamma^3\beta},
\eeq
where we used $\partial_t(r\Bbg)_\xi=c\,d(r\Bbg)/dr=-2c\Bbg$ and $d\gamma=\gamma^3\beta\,d\beta$. Then, substituting \Eq~(\ref{eq:E_gamma}) into \Eq~(\ref{eq:en2}), we find
\beq
\label{eq:evol_gamma}
   \left[\frac{2\sigbg}{\gamma^3(1-\beta)^3} +\frac{1}{1-\beta}\right]\partial_t\gamma
   +\partial_\xi\gamma = \frac{4c\,\sigbg \beta^2}{r\,(1-\beta)^2}. 
\eeq
Note that the term $(1-\beta)^{-1}$ in the bracket (brought by $\partial_t\gamma$ on the r.h.s. of \Eq~(\ref{eq:en2})) is negligible only if $2\gamma^3\ll \sigbg$. This condition corresponds to $(\partial_t U_{\rm pl})_\xi\ll (\partial_t U_{\rm w})_\xi$ where $U_{\rm pl}=\gamma n\me c^2$ and $U_{\rm w}=E^2/4\pi$.

A more convenient variable related to $\beta$ is the compression of proper density $\trho=\rho/\gamma$, 
\beq
\label{eq:c}
    \c \equiv \frac{\trho}{\rhobg} = \frac{\tB}{\Bbg} =\frac{\sigma}{\sigbg} = 
     \sqrt{\frac{1+\beta}{1-\beta}},
\eeq
where $\trho$ and $\tB$ are measured in the fluid rest frame, and $\sigma\equiv \tB^2/4\pi\trho c^2$. Note that $-1<\beta<1$ corresponds to $-\infty <\c<\infty$.
\Eq~(\ref{eq:evol_gamma}) rewritten in terms of $\c$ becomes 
\beq
\label{eq:cold_wave}
      \left( 2\sigbg \c^3 + \frac{1}{2} \right) \partial_t \c
  +  \partial_\xi \c 
  =  \frac{2c}{r}\,\sigbg\c^2(\c^2-1).
\eeq
Here, in the coefficient of $\partial_t \c$ we neglected $\c^2/2\ll 2\sigbg\c^3$, using
\beq
  2\sigbg \c^3 + \frac{1+\c^2}{2} = 2\sigbg \c^3\left[1+{\cal O}\left(\frac{1}{\sigbg\c} \right) \right] + \frac{1}{2}.
\eeq

The equatorial wave evolution will be found if we solve \Eq~(\ref{eq:cold_wave}) for $\c(t,\xi)$. This first-order partial differential equation is linear in the derivatives $\partial_t\c$ and $\partial_\xi\c$, and can be solved using the method of characteristics.

\subsection{Characteristics}

One can rewrite \Eq~(\ref{eq:cold_wave}) as
\beq
 \label{eq:evol}
  \left.\frac{d\c}{dt}\right|_{C^+} = \frac{c\,(\c-\c^{-1})}{ r\,[1 + (4\sigbg\c^3)^{-1}]},
\eeq
where the derivative $d\c/dt$ is taken along curves $C^+$ (characteristics) defined by 
\beq
\label{eq:C+}
   \frac{d\xi_+}{dt} = \frac{2}{4\sigbg\c^3+1}.
\eeq
The characteristics $\xi_+(t)$ can also be described by their radial speed,
\beq
\label{eq:beta+}
   \beta_+ = \frac{1}{c}\,\frac{dr_+}{dt} = 1-\frac{d\xi_+}{dt} = \frac{4\sigbg\c^3-1}{4\sigbg\c^3+1}.
\eeq

In an accompanying paper we give an alternative derivation of the wave evolution equation generalized to relativistically hot plasmas, and  discuss two families of MHD characteristics $C^\pm$ propagating with radial speeds\footnote{$C^-$ did not explicitly appear in our derivation above. For short waves, integration of MHD equations along $C^-$ is just another way to get the relation between $n$, $B$, and $\beta$ (\Eq~\ref{eq:continuity_short}), and the wave evolution problem is reduced to integration along $C^+$.} 
\beq
   \beta_\pm = \frac{\beta\pm\bs}{1\pm\beta\bs}.
\eeq
Here, $\bs$ has the meaning of wave speed relative to the plasma (the ``magnetosonic speed''). In a magnetically dominated plasma $\bs$ is close to unity. In particular, for a cold plasma, 
\beq
   \bs^2=\frac{\sigma}{1+\sigma}=\frac{\c\sigbg}{1+\c\sigbg}.
\eeq

The FFE limit corresponds to $\sigbg\rightarrow\infty$ and $\bs\rightarrow 1$. In this limit, \Eqs~(\ref{eq:evol}) and (\ref{eq:C+}) simplify to $d\c/d\ln r=\c-\c^{-1}$ and $dr_+/dt=c$ ($\Leftrightarrow d\xi_+/dt=0$), and give the solution
\beq
\label{eq:cC+}
     \c=\sqrt{1+2Kr^2}, 
\eeq
where $K=const$ along $C^+$. The corresponding solution for $E$ is 
\beq
\label{eq:EK}
    E=\frac{\mu K}{r} \qquad {\rm (FFE)}.
\eeq

\subsection{Shock formation}
\label{shock_formation}

Significant MHD corrections to FFE arise if $\sigma=\c\sigbg$ drops, i.e. if the plasma experiences strong expansion in the wave, $\c\ll 1$. This occurs where the plasma is accelerated as $E^2$ approaches $B^2$. Note that strong expansion also implies strong adiabatic cooling. Thus, shock formation is expected in the part of the wave with large $\gamma$, small $\c$, and a reduced temperature. The process of MHD shock formation is illustrated in Figure~3. The shock occurs where the characteristics cross, bringing different values of $\beta$ to the same location and thus creating a discontinuity.

\begin{figure}[t]
\begin{center}
\includegraphics[width=0.4\textwidth]{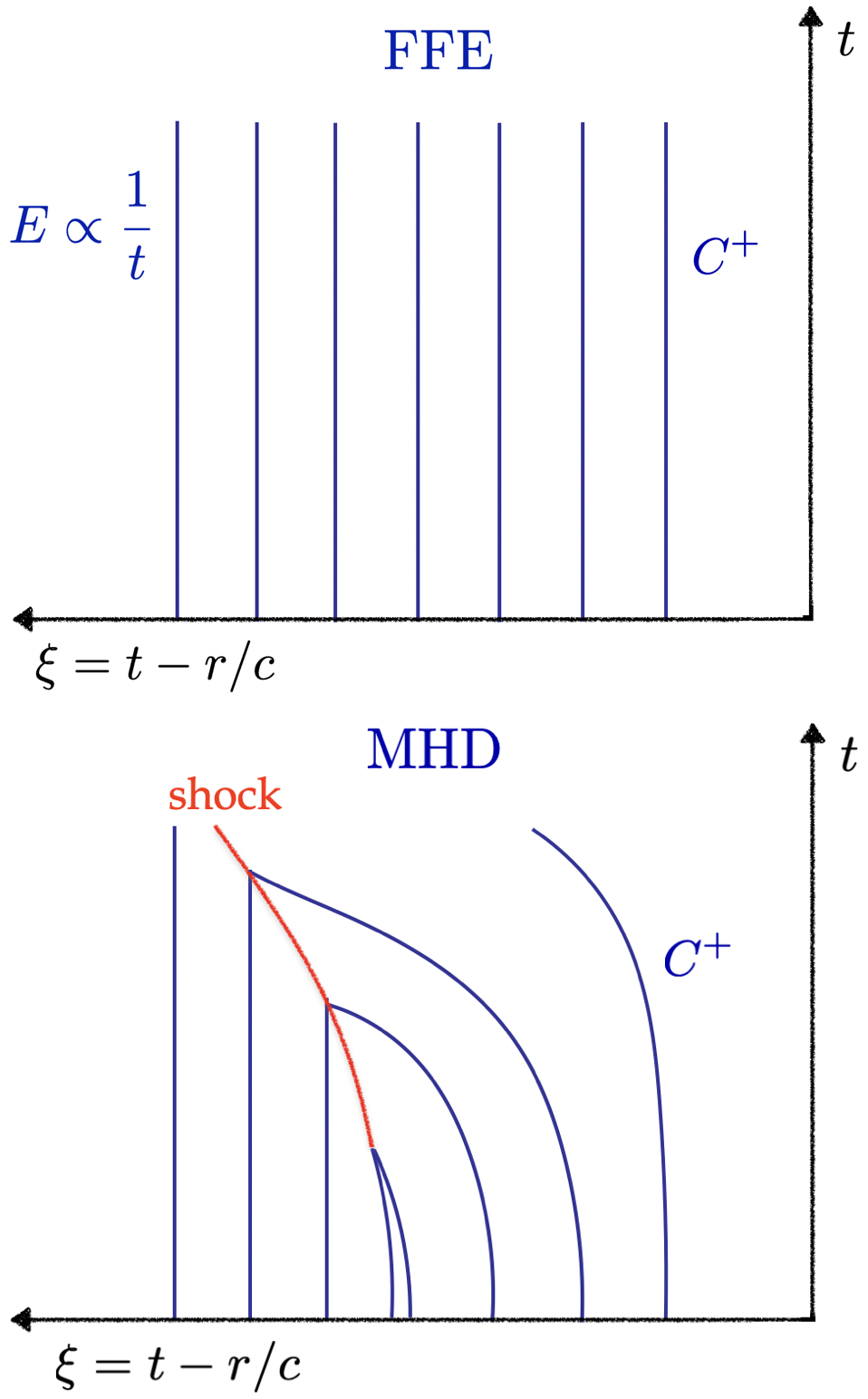} 
\caption{Schematic illustration of the flow of characteristics $C^+$ in FFE and MHD, shown in the $t$-$\xi$ plane. Each $C^+$ has its initial position $\xiin$ and carries a value of $K(\xiin)=rE/\mu$ determined by the initial wave profile. In the FFE limit, the characteristics are vertical straight lines; they propagate with speed $\beta_+=1$, which corresponds to $d\xi_+/dt=0$. The MHD correction to FFE implies $d\xi_+/dt=1-\beta_+\neq 0$, so the $C^+$ characteristics are no longer static in $\xi$; they become significantly bent in the region where $E^2$ approaches $B^2$, leading to the formation of a shock (red curve). }
\end{center}
\label{fig:C+}
 \end{figure}

We consider waves emitted at sufficiently small radii where $\Bbg$ far exceeds  the wave electric field $E$, and the plasma oscillates with small $|\beta|\ll 1$, which implies a negligible change in plasma density, $|\c-1| \ll 1$. At the small radii, characteristics propagate with speed $\beta_+=1-{\cal O}(\sigbg^{-1})$ and FFE is an excellent approximation. In particular, the wave initially propagates with negligible distortion: each $C^+$ characteristic satisfies $d\xi_+/dt=0$, i.e. keeps a constant coordinate $\xi=t-r/c=\xiin$, and one can define an initial (undeformed) profile of $E(\xiin)$. More precisely, the profile of $rE(\xi)$ is static at small $r$ while the normalization of $E(\xi)$ is decreasing, $E\propto r^{-1}$. 

This initial profile is the only parameter of the problem besides the magnetospheric parameters $\mu=r^3\Bbg$ and $D=r^3\sigbg$. It is conveniently described by 
\beq 
\label{eq:K_xiin}
   K(\xiin)\equiv \frac{rE}{\mu} \quad ({\rm set~at~small~} r).
\eeq
We will calculate the wave evolution by tracking the $C^+$ characteristics, each described by its initial position $\xiin$ and $K(\xiin)$. Note that $K(\xiin)$ may be positive or negative. Characteristics with $K<0$ will develop $\c\ll 1$, leading to monster shock formation.

It is easy to see that even an arbitrarily high $\sigbg\rightarrow \infty$ does not save the wave from breaking. Indeed, consider a wave with $E$ oscillating between $\pm \Em$. As the wave expands from small radii, where $\Em(r)\ll \Bbg(r)$, the ratio $\Em/\Bbg\propto r^2$ grows and eventually approaches $1/2$ at some radius $\Rm$. At this moment, the minimum $E=-\Em=-\Bbg/2$ and
\beq
    E=-B, \quad \beta=-1, \quad \gamma\rightarrow\infty, \quad   \c\rightarrow 0.
\eeq
An arbitrarily high $\sigbg$ does not prevent the decrease of $\sigma=\c\sigbg$, which reduces the speed of $C^+$ characteristics, $dr_+/dt<c$. The characteristics become bent and eventually collide, forming a shock (Figure~3). 

A formal proof of shock formation is provided by the solution of the coupled \Eqs~(\ref{eq:evol}) and (\ref{eq:C+}) for $\c(t)$ and $\xi_+(t)$ or $r_+(t)=c(t-\xi_+(t))$ (Appendix~\ref{cold_wave}). Numerical examples will be given in \Sect~\ref{numerical}.


\section{Method for tracking shocks}
\label{shock}

\subsection{Shock strength}

The plasma speed $\beta$ is discontinuous at the MHD shock because the upstream and downstream characteristics bring to the shock different values of $\beta$: $\betau\neq \betad$ (subscripts ``u'' and ``d'' refer to the immediate upstream and downstream of the shock).
The corresponding jump in the proper density characterizes the shock strength,
\beq 
    q\equiv\frac{\trhod}{\trhod}=\frac{\cd}{\cu}=\frac{\gd(1+\betad)}{\gu(1+\betau)}.
\eeq
It is related to the Lorentz factor of the upstream plasma relative to the downstream, $\Grel=\gu\gd(1-\betau\betad)$. A similar expression for $\Grel$ holds in the shock rest frame $\KF'$. To distinguish between different frames, dynamical quantities measured in the shock frame will be denoted with a prime, and quantities measured in the drift frame $\tilde{\KF}$ (in which $\tilde{E}=0$) are denoted with tilde.
Using $\gu',\gd'\gg 1$, one can write the continuity of mass flux as $\trhou\gus\approx\trhod\gds$, and find
\beq
  \Grel \approx \frac{1}{2} \left( \frac{\gu'}{\gd'}+\frac{\gd'}{\gu'} \right)
  \approx\frac{1}{2} \left( q+\frac{1}{q} \right). 
\eeq

\subsection{Shock speed}

We can track the evolution of waves with shocks if we know the shock Lorentz factor relative to the downstream plasma $\gamma_{\rm sh|d}$. The standard jump conditions for perpendicular relativistic magnetized shocks give $\gamma_{\rm sh|d}\approx\sqrt{\sigu}\gg 1$. The shock is collisionless and mediated by Larmor rotation, so the jump normally occurs on the Larmor scale, as verified by detailed kinetic simulations \citep{Sironi21}. Dissipation of the upstream kinetic energy occurs through gyration and thermalization enabled by the instability of coherent gyration in the downstream.

However, the standard picture of Larmor-mediated shocks can fail for the monster shocks, where particles experience extremely fast radiative losses. The ultra-relativistic plasma flow entering the shock is subject to radiation reaction in response to curvature in the particle trajectory. The upstream flow may experience a strong radiative drag before it completes one Larmor rotation and joins the downstream.

\begin{figure}[t]
\vspace*{-0.8cm}
\begin{center}
\includegraphics[width=0.46\textwidth]{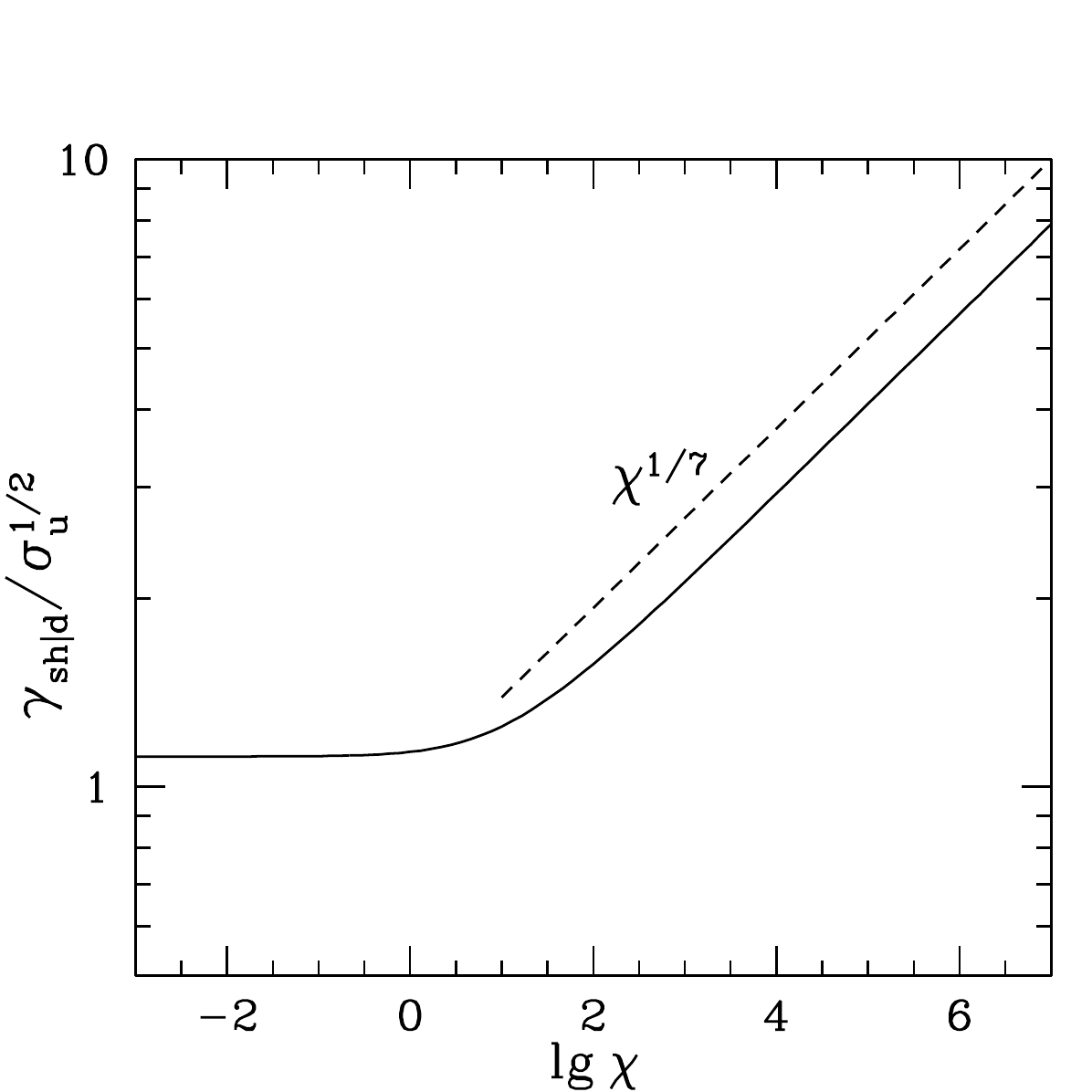}
\end{center}
\caption{Lorentz factor of the shock relative to the downstream plasma as a function of the radiative parameter $\chi$ (solid curve, calculated as explained in Appendix~\ref{shock_structure}). Dashed line shows the analytical result at $\chi\gg 1$ (\Eq~\ref{eq:jump}).
}
\label{fig:jump}
 \end{figure}

The resulting jump conditions can be evaluated by going beyond the MHD description and examining microphysics of the shock transition on scales smaller than the particle Larmor radius. This is done in Appendix~\ref{shock_structure}, where we calculate the structure of the flow across the jump numerically and also derive an approximate solution analytically.  The result is shown in Figure~\ref{fig:jump}, and the jump condition may be stated as
\beq
\label{eq:jump}
   \gamma_{\rm sh|d}\approx (1+\chi)^{1/7} \sqrt{\sigu}, 
\eeq
which smoothly matches the results at $\chi\ll 1$ and $\chi\gg 1$. The parameter $\chi$ is defined in \Eq~(\ref{eq:chi}) in terms of the upstream Lorentz factor $\gu'$ and magnetic field $\Bu'$ measured in the shock frame. Using $\Bu'=\gus\tBu$ and $\gus=q\gds=q\gamma_{\rm sh|d}$, one can express it as
\beq
\label{eq:chi1}
  \chi = \frac{\sT\tilde{B}_{\rm u} q^3 \gamma_{\rm sh|d}^3}{4\pi e\,\sigu^{3/2}},
\eeq
where $\tilde{B}_{\rm u}=\cu\Bbg$ and $\sigu=\cu\sigbg$.
\Eqs~(\ref{eq:jump}) and (\ref{eq:chi1}) can be solved for $\gamma_{\rm sh|d}$ and $\chi$, which gives 
\beq
\label{eq:jump2}
     \chi^{4/7} \approx \frac{\sT \Bbg \cd^3}{4\pi e \cu^2}.
\eeq
This expression is found for $\chi\gg 1$, and the solution for $\chi<1$ is not needed as its value makes no difference for $\gamma_{\rm sh|d}\approx (1+\chi)^{1/7}\sqrt{\sigu}\approx \sqrt{\sigu}$. 

The shock Lorentz factor in the lab frame is 
\beq 
  \gsh= \gamma_{\rm sh|d}\gd(1+\beta_{\rm sh|d}\betad)
  \approx \gamma_{\rm sh|d}\cd,
\eeq
where we used $1+\beta_{\rm sh|d}\betad\approx 1+\betad$, because $\gamma_{\rm sh|d}\gg \gd$. 
The shock motion in the $\xi$-coordinate is described by $d\xish/dt=1-\bsh\approx (2\gsh^2)^{-1}$. Thus, we find
\begin{eqnarray}
\label{eq:shock_speed}
    \frac{d\xish}{dt}
    \approx \frac{1}{2\gsh^2}
    \approx \left\{\begin{array}{cl}
          \displaystyle{\left(\frac{\pi e}{\sT\Bbg \sigbg^2 \cd^7}\right)^{1/2}} & \quad \chi> 1 \\
          (2\sigbg\cu\cd^2)^{-1} & \quad \chi<1
                         \end{array}\right.
\end{eqnarray} 

There is one caveat in the derivation of this result: we neglected that the shock can generate an electromagnetic precursor, which interacts with the upstream flow and may reduce its Lorentz factor $\gu$ (\Sect~\ref{precursor}). Note however that $d\xish/dt$  is independent of $\cu\approx (2\gu)^{-1}$ when $\chi\gg 1$. Furthermore,  the shock speed will be dynamically regulated to a value independent of $\cu$ and $\cd$ when $\cd\ll 1$, as explained in \Sect~\ref{strength}.

\subsection{Tracking waves with shocks}

The MHD wave evolution is controlled by the flow of $C^+$ characteristics in coordinate $\xi=t-r/c$. This flow is described by \Eqs~(\ref{eq:evol}) and (\ref{eq:C+}). They determine both the shape of characteristics and the values of $\c$, $\beta$, and $E$ on each $C^+$. 

After caustic formation, the born shock separates the $C^+$ flow into two regions: upstream and downstream (ahead and behind the shock). The characteristics propagate with different speeds in these regions and collide at the shock. The colliding characteristics $\Cu^+$ and $\Cd^+$ are terminated and disappear from the wave evolution problem. The location of $\Cu^+$-$\Cd^+$ collision moves with speed $\bsh$ determined by \Eq~(\ref{eq:shock_speed}). The shock always propagates faster than $\Cu^+$ and slower than $\Cd^+$: $\betau^+<\bsh<\betad^+$. 

\Eqs~(\ref{eq:evol}), (\ref{eq:C+}), and (\ref{eq:shock_speed}) give a closed description for MHD waves with shocks. The cold approximation $\gs^2=\sigma$ used in \Eqs~(\ref{eq:C+}) turns out sufficient for the kHz waves studied in this paper. The accurate $\gs<\infty$ is important in the pre-shock region where $\c$ drops so much that the $C^+$ flow experiences significant deformation $d\xi_+/dt$ (Figure~3). This region is also coldest (due to adiabatic cooling accompanying the drop in $\c$), and so $\gs^2=\sigma$. In the post-shock region $\xi>\xish$, fast radiative losses also allow one to use $\gs^2\approx\sigma$. Here, the evolution is anyway simple, not sensitive to the precise $\gs$: the post-shock $\gs$ is so high that $C^+$ remain practically static in the $\xi$ coordinate during the main shock dissipation phase $r\simlt 3\Rm$.

\newpage

\subsection{Numerical implementation}

The described method for calculating the wave evolution is easily implemented in a numerical simulation. When launching a wave, we set up an initially uniform grid in $\xiin$ of size $N_+$, and then use the $N_+$ characteristics to track the wave evolution in the $\xi$ coordinate. In the simulation presented below we used $N_+=10^5$. At each timestep $dt$, the displacement $d\xi_+$ of each characteristic and the change of its $\c$ are determined by \Eqs~(\ref{eq:C+}) and (\ref{eq:evol}). After each timestep, the code examines the updated positions or the characteristics and checks for their crossing to detect shock formation. Once the shock is born, the code begins to track its motion according to \Eq~(\ref{eq:shock_speed}) and also check at each timestep which characteristics terminate at the shock. 

The simulation thus follows the entire evolution of the wave, from its initial deformation-free propagation at $r<\Rm$ to shock formation at $r\approx\Rm$ to subsequent evolution with the embedded shock. We use an adaptive timestep to resolve fast changes in MHD quantities that occur  near $\Rm$. Note also that the density of characteristics drops ahead of the shock, where $\c$ is lowest and $d\xi_+/dt$ is highest. To maintain sufficient spatial resolution we use adaptive mesh refinement in $\xiin$ without changing the total number $N_+$ of active (not terminated) characteristics. This is achieved by launching new characteristics in the region of low resolution while discarding the characteristics terminated at the shock.

The calculated $\xi_+(t)$ and $\c(t)$ along each $C^+$ determine the wave profile $\c(\xi)$ at any time $t$. Thus, we find the evolution of the wave profile.


\section{Sample numerical models}
\label{numerical}

As a concrete example, consider a wave with the initial profile $E(\xiin)=\Em\sin(\omega\xiin)$. Recall that $E\propto r^{-1}$ along each characteristic $C^+$ until it develops an extremely low compression factor $\c$ where $B^2-E^2$ approaches zero. The initial profile of $E$ sets the parameter $K=rE/\mu$ on each $C^+$,
\beq
\label{eq:sine}
   K(\xiin)=\Km \sin(\omega\xiin), \qquad \Km=\frac{r\Em(r)}{\mu}.
\eeq

\begin{figure*}[t]
\hspace*{-0.5cm}
\includegraphics[width=0.6\textwidth]{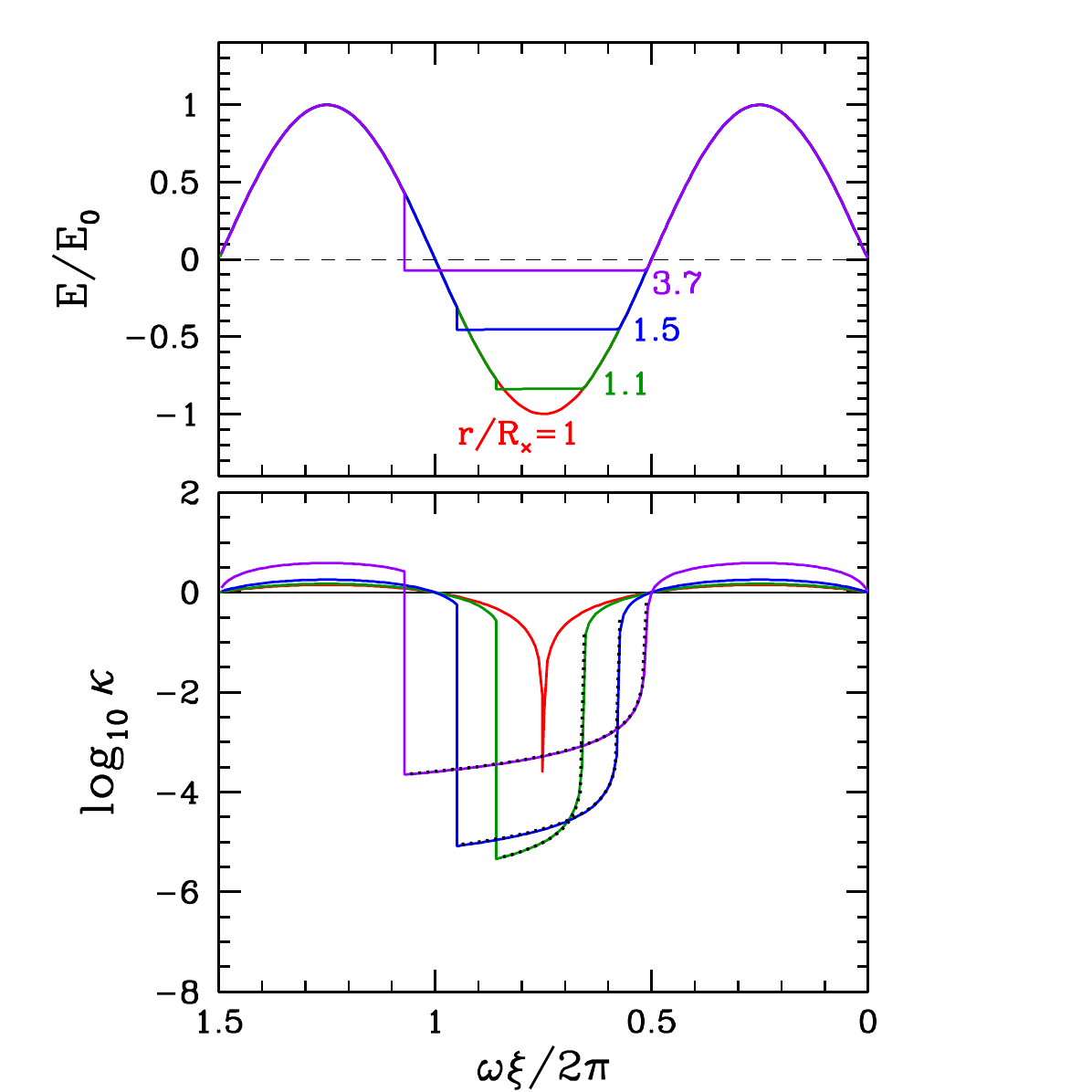} 
\hspace*{-2cm}
\includegraphics[width=0.6\textwidth]{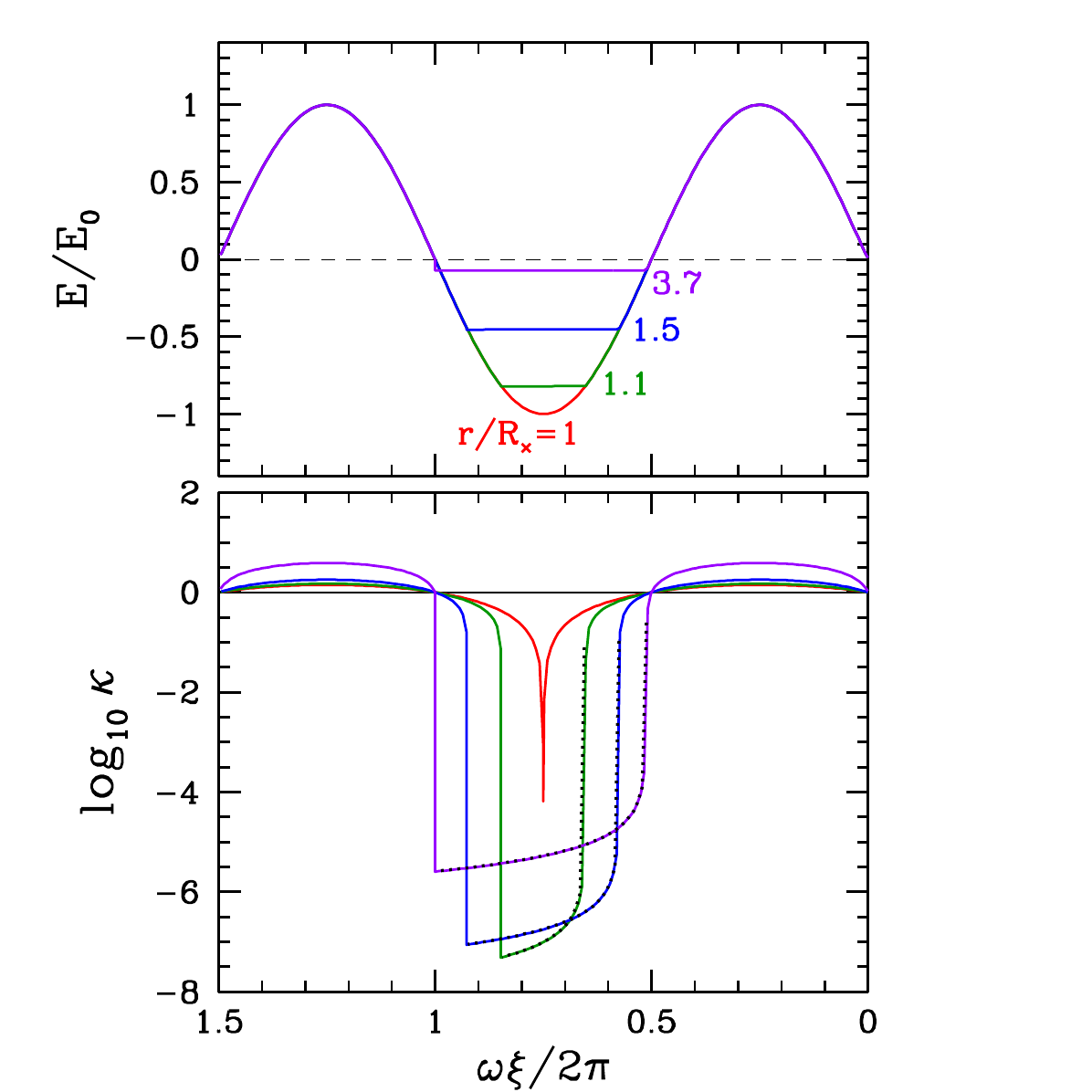} 
\vspace*{-3mm}
\label{fig:kHz}
\caption{{\it Left:} Evolution of the wave profile $E(\xi)$ and $\c(\xi)$, where $\xi\equiv t-r/c$. The wave has frequency $\nu=\omega/2\pi=10\,$kHz and initial power $L=10^{43}\,$erg/s; the magnetosphere has magnetic dipole moment $\mu=10^{33}$\,G\,cm$^3$ and density parameter $\N=10^{39}$. Five snapshots are shown, when the wave packet reaches $r/\Rm=0.9$ (black), 1 (red), 1.1 (green), 1.5 (blue), and 3.7 (magenta). Electric field $E$ is normalized to $\Em$ that would be the wave amplitude if it propagated in vacuum. The plasma Lorentz factor $\gamma$ is related to the proper compression $\c$ by $\gamma=(1+\c^2)/2\c$. Black dotted curves show the analytical result for $\c(\xi)$ (\Eq~\ref{eq:c_profile}). The simulation neglected the shock precursor effect, which can reduce $\gamma$ in the interval $3\pi/2\omega<\xi<\xish$ (\Sect~\ref{precursor}).
{\it Right:} Same wave but now launched into the magnetosphere with $\N=10^{37}$. 
Here, $\c$ becomes extremely small, breaking the MHD description and transitioning to a two-fluid regime. We argue in \Sect~\ref{MHD_validity} that the two-fluid calculation will likely give the same evolution of $E(\xi)$ and $\c(\xi)$ as found in the MHD model.}
 \end{figure*}

The wave trough $E=-\Em$ approaches $-\Bbg/2$ (so $E^2$ approaches $B^2$) at radius $\Rm$. It is related to $\Km$ by
\beq
\label{eq:Rm}
     \Rm\approx\frac{1}{\sqrt{2\Km}}.
\eeq
Radius $\Rm$ is also related to the isotropic equivalent of the wave power $L\approx c r^2 \Em^2/2$ and the magnetic dipole moment of the star $\mu$:
\beq 
  \Rm=\left(\frac{c\mu^2}{8L}\right)^{1/4}
  \approx 1.4\times 10^8\, \frac{\mu_{33}^{1/2}}{L_{43}^{1/4}}{\rm ~cm}.
\eeq
The background magnetization parameter at $\Rm$ is determined by \Eq~(\ref{eq:sigbg}),
\beq
   \sigm\equiv\sigbg(\Rm)
   \approx 3.6\times 10^{9}\,\frac{\mu_{33}^{1/2}L_{43}^{3/4}}{\N_{37}}.
\eeq
Our sample models assume the plasma density parameter $\N=10^{39}$ and $10^{37}$ (\Sect~\ref{bg}). In both models, the magnetosphere has the dipole moment $\mu=10^{33}$\,G\,cm$^3$. The wave has frequency $\nu=\omega/2\pi=10\,$kHz and initial power $L=10^{43}\,$erg/s. The simulation results are shown in Figure~5 and may be summarized as follows.

\begin{figure}[t]
\hspace*{1mm}
\includegraphics[width=0.46\textwidth]{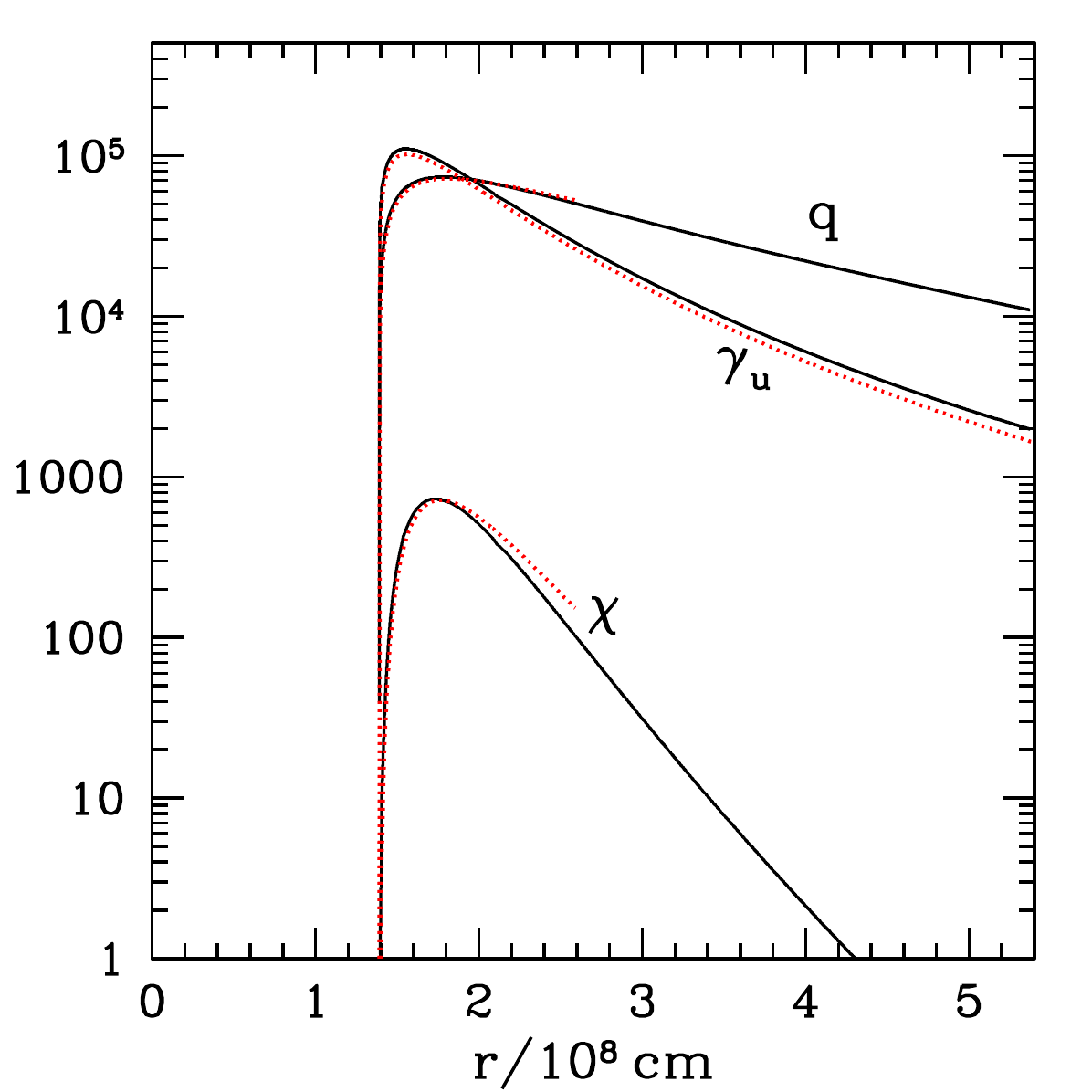} 
\caption{
Evolution of shock parameters in the simulation with $\N=10^{39}$:  upstream Lorentz factor $\gu$, shock compression factor $q\equiv\cd/\cu$, and  radiative parameter $\chi$. Red dotted curves show the analytical results for $\gu(r)$, $q(r)$, $\chi(r)$ derived in \Sect~\ref{analytical} (\Eqs~(\ref{eq:cu}), (\ref{eq:q}),  (\ref{eq:chi2})). The simulation neglected the precursor effect (\Sect~\ref{precursor}).}
\label{fig:kHz_q}
 \end{figure}

The wave travels with no distortion until it comes to $\Rm$ where $E^2$ nearly reaches $B^2$. The plasma at this point develops an ultra-relativistic drift with $u\approx -(1/2)\sqrt{\Bbg/(B-|E|)}$, leading to immediate shock formation. In particular, in the model with $\N=10^{39}$ the plasma four-velocity upstream of the shock reaches $\uu\sim -10^5$, and the shock compression factor $q=\cd/\cu$ exceeds $10^4$. The shock has a large radiative parameter $\chi\gg 1$(Figure~\ref{fig:kHz_q}). An analytical derivation reproducing the numerical results will be given in \Sect~\ref{analytical}. We will show there that the Lorentz factor of the accelerated plasma scales as $\gamma\propto L/\N\nu$, and $\N\sim 10^{37}$ gives so high $\gamma$ that the MHD description breaks and the plasma dynamics needs a two-fluid description. 

As one can see in Figure~5, the wave profile $E(\xi)$ develops a plateau of $E\approx -\Bbg/2$, which corresponds to $E^2\approx B^2$. The small variation $\delta E$ along the plateau implies that the plateau is formed by stretching a small interval $\delta \xi$ of the initial profile $E(\xi)$. The stretching is clearly demonstrated by the $C^+$ flow. The part of the wave profile where $E^2$ approaches $B^2$ develops a low $\c^2=(B-|E|)/\Bbg$, and here $d\xi_+/dt$ steeply increases (\Eq~\ref{eq:C+}). As a result, characteristics with $E^2\approx B^2$ swiftly ``fall'' toward the shock, with acceleration, like a waterfall. This effect protects the MHD wave from breaking the condition $E^2<B^2$. Where $\c$ becomes so low that $\c^3\sigbg\ll 1$, the $C^+$ motion in $\xi$ saturates at $d\xi_+/dt\approx 2$. The fast, large displacement of $C^+$ in $\xi$ creates a nearly perfect plateau of $E\approx -B\approx -\Bbg/2$, with a small difference $B-|E|\approx \c^2\Bbg$.

As the wave propagates to larger radii and $\Bbg\propto r^{-3}$ decreases, the plateau level $E=-\Bbg/2$ moves up and away from the original minimum of the sine wave ($E=-\Em$), and hence its width grows, so the plateau is forced to occupy an increasing part of the wave profile. At radii $r\gg\Rm$, where $\Bbg/2\Em=\Rm^2/r^2\ll 1$, the plateau approaches $E\approx 0$, i.e. the part of the wave with $E<0$ becomes erased. As a result, the wave loses about half of its original energy.\footnote{Maximum dissipation fraction of the shock is $\sigu^{-1}\sim c/\omega\Rm\ll 1$. This small factor is compensated by the large ratio of the traveled distance $\sim \Rm$ to $c/\omega$, resulting in strong dissipation.} Waves with multiple oscillations lose even more than half: one can see that the shock formed in the first oscillation at $\xi\approx 3\pi/2\omega$ eventually enters the second oscillation $\xi>2\pi/\omega$ before stalling there, so the final plateau of $E=0$ occupies slightly more than half of the oscillation.

The plasma speed $\beta=E/B=E/(\Bbg+E)$ is positive where $E>0$ and reaches maximum at the wave crest. This maximum $\beta$ approaches unity at $r\gg\Rm$. Here, the plasma speed relative to the wave, $1-\beta\approx \Bbg/E$, becomes small, increasing the time it takes the plasma to cross the first half of the wave oscillation,
\beq
\label{eq:tcross}
    \tcross=\int_0^{\pi/\omega} \frac{d\xi}{1-\beta}
    \approx \frac{r^2}{\omega\Rm^2} \quad (r\gg\Rm).
\eeq
The short-wave approximation $\tcross\ll r/c$ holds throughout our simulation. The simulation did not follow how the plasma eventually becomes trapped in the wave, $\tcross> r/c$; this occurs later, when the wave propagates to larger radii. 

Besides the monster shock launched at $\xi\approx 3\pi/2\omega$ at radius $\Rm$, the simulation shows gradual steepening of the wave at the leading edge $\xi=0$ (and at $\xi\approx 2\pi/\omega$). At radius $\RF\approx 10\Rm$, the wave launches a forward shock as shown in \Sect~\ref{FS}.


\section{Analytical description}
\label{analytical}

The wave evolution demonstrated by the numerical simulation may also be derived analytically. Two key dimensionless parameters of the problem are
\begin{eqnarray}
    \zeta &\equiv& \frac{\omega\Rm}{c \,\sigm}=\pi^2\,\frac{\me c^2\N \nu}{L}
    \approx 8\times 10^{-8}\,\N_{37}L_{43}^{-1}\nu_4, \;\;\\
   \eta &\equiv& \frac{\sT\Bm}{8\pi e} = \frac{r_e\omegam}{3c}
   \approx 2\times 10^{-8}\,\mu_{33}^{-1/2}L_{43}^{3/4}, 
\end{eqnarray}
where $\Bm\equiv\Bbg(\Rm)$ and $\omegam\equiv e\Bm/\me c$.

 \subsection{Shock formation: caustic in the $C^+$ flow}
 \label{caustic_harmonic}

Each $C^+$ starts at small radii with $\xi_+=\xiin$, and one can think of $\xiin$ as a Lagrangian coordinate in the flow of characteristics $\xiin\rightarrow\xi_+(\xiin,t)$ with initial condition $\xi_+(\xiin,0)=\xiin$. The $C^+$ flow satisfies the relation~(\ref{eq:xiC+}) found from \Eqs~(\ref{eq:evol}) and (\ref{eq:C+}):
\beq
\label{eq:flow_xi+}
    \xi_+ = \xiin+\frac{(\c-1)^2}{8 cD \K^2\c} = \xiin + \frac{\gamma-1}{4cD\K^2},
\eeq
where $D=\sigbg r^3=\sigm\Rm^3$. The caustic in the $C^+$ flow (birth of the shock) appears where $(\partial\xi_+/\partial\xiin)_t$ vanishes. It means that characteristics carrying different values of $\K$ begin to cross, creating a discontinuity.
This happens at some $\xiin=\xiinc$ and time $t=\tc$, which may be derived analytically (Appendix~\ref{cold_wave}). We also find the plasma compression factor at the caustic, $\cc$. 

In the leading order of the small parameters $\zeta$ and $(\Rm\omega/c)^{-1}$ we obtain
\beq
\label{eq:caustic}
    \cc=\frac{\sqrt{\zeta}}{24^{1/4}}, 
    \qquad \cos(\omega\xiinc)=\frac{4\sqrt{\zeta}}{24^{3/4}} - \frac{c}{\Rm\omega},
\eeq
\beq
\label{eq:tc}
 \frac{c\tc}{\Rm} - 1 = -\frac{\zeta}{3\sqrt{24}}-\frac{8\sqrt{\zeta}}{24^{3/4}}\frac{c}{\Rm\omega}+\frac{c^2}{4\Rm^2\omega^2}.
\eeq
Powerful kHz waves have $\sqrt{\zeta}\ll c/\Rm\omega$. In this regime, $C^+$ characteristics turn back toward the star ($\beta_+<0$) before forming the caustic. The caustic occurs at $\xi_c$ near $3\pi/2\omega$:
\beq
\label{eq:xic}
   \omega\xi_c\approx \frac{3\pi}{2} + \frac{16\sqrt{\zeta}}{24^{3/4}} - \frac{c}{\Rm\omega}.
\eeq

\subsection{Plasma motion upstream of the shock}
\label{gu}

As one can see in Figure~5, the plasma develops a very low compression factor $\kappa$ ahead of the shock, a result of huge plasma acceleration toward the star. The acceleration occurs along the plateau of $E\approx -\Bbg/2$, and the four-velocity $u$ of the plasma drift in the wave is controlled by how close $B^2-E^2$ approaches zero:
\beq 
  u\approx -\gamma \approx -\frac{B}{\sqrt{B^2-E^2}}
  \approx -\frac{1}{2} \sqrt{\frac{\Bbg}{\Bbg + 2E}}.
\eeq
The plasma acceleration is accompanied by its expansion by the factor of $\c^{-1}\approx 2\gamma$, as follows from the relation
\beq
   \frac{\c^2+1}{2\c}=\gamma, \qquad \frac{\c^2-1}{2\c}=u.
\eeq 
It is the drop of $\c$ that boosts the motion of $C^+$ characteristics in the $\xi$-coordinate, deforming the wave profile. 
The relation between $\c$ and the displacement of characteristics $\xi_+-\xiin$ is given by \Eq~(\ref{eq:flow_xi+}). Substituting $\K=\Km\sin(\omega\xiin)$ and $\Km^{-1}=2\Rm^2$, we find 
\beq
\label{eq:xi_kappa}
   \xi_+-\xiin = \frac{(1-\c)^2\Rm}{2c\, \c\, \sigm \sin^2(\omega\xiin)}.
\eeq
One can use this relation to evaluate $\cu$  on the characteristic $\Cu^+$ that reaches the shock at time $t$: $\xiu_+=\xish$. We are interested in waves that develop $\cu\ll 1$, and then \Eq~(\ref{eq:xi_kappa}) gives 
\beq
\label{eq:cu_}
  \cu\approx \frac{\zeta}{2\,\omega\Delta\xi\sin^2(\omega\xiinu)}, 
\eeq
where $\Delta\xi\equiv\xiu_+-\xiinu$ is the displacement of $\Cu^+$.

For the nascent shock at the caustic we find $\omega\Delta\xi \approx 12\sqrt{\zeta}/24^{3/4}$, which corresponds to $\cu=\cc$ (\Sect~\ref{caustic_harmonic}). Later, $\Delta\xi(t)$ approximately equals the width of the plateau,
\beq
   E\approx \Ep=-\frac{\Bbg}{2} \qquad ({\rm plateau}), 
\eeq
along which $\c$ drops to $\cu$ (Figure~5). All $C^+$ populating the plateau have approximately the same $E$ and so nearly the same $\xiin$ --- the plateau forms by the huge stretching (by the factor of $\sim \c^{-1}$) of an initially small interval in the original wave profile $\delta\xiin$. This stretching occurs because the characteristics ``fall'' onto the shock from $\xiin$ where $E$ approaches $-\Bbg/2$, i.e. $\sin(\omega\xiin)\approx -\Bbg/2\Em$. 

The plateau begins and ends where $\sin(\omega\xi)\approx-\Bbg/2\Em\approx -\Rm^2/r^2$, so its width is $\omega\Delta\xi\approx\pi - 2\arcsin(\Rm^2/r^2)$. Then, from \Eq~(\ref{eq:cu_}) we find the plasma Lorentz factor just upstream of the shock,
\beq
\label{eq:cu}
  \gu\approx \frac{1}{2\cu}\approx
  \frac{1}{\zeta x^4}\left(\pi - 2\arcsin\frac{1}{x^2}\right), \quad x\equiv\frac{r}{\Rm}.
\eeq
This expression accurately reproduces the simulation results (Figure~\ref{fig:kHz_q}). The maximum $\gu\approx \zeta^{-1}$ is reached at $r/\Rm\approx(1+1/\pi)^{1/2}\approx 1.15$.

The same $\gu$ can be found from energy conservation, as the flow is accelerated at the expense of the electromagnetic wave energy $\E=\int c\,r^2 E^2\,d\xi$ (isotropic equivalent). The profile $E(\xi)$ includes the plateau part ($E\approx\Ep=-\Bbg/2$) and the part where $E(\xi)$ is practically unchanged from the vacuum propagation $E(\xi)\approx \Em\sin(\omega\xi)$ (Figure~5). The plateau has the width $\Delta\xi$ and ends at the shock $\xish$ with a jump $[E^2]=\Ed^2-\Eu^2$, where $\Eu\approx\Ep$. During time $dt\approx dr/c$, the ``area'' associated with the integral $\int r^2E^2d\xi $ changes: (1) the rise of the plateau erases a horizontal stripe $|d(r^2\Ep^2)|\,\Delta\xi$ and (2) the motion of the shock erases a vertical stripe $r^2[E^2]\,d\xish$. So, the wave loses energy 
\begin{eqnarray}
  d\Elost \approx -c \,d(r^2\Ep^2)\, \Delta\xi + cr^2[E^2]\,d\xish.
  \label{eq:dEdiss2}
\end{eqnarray}
The energy lost at $\xi<\xish$ (the first term on the r.h.s.) goes into plasma acceleration upstream of the shock, 
\beq
\label{eq:gu2}
   \gu \me c^2 d\N \approx 4 c\,r^2 \Ep^2\, d\ln r\, \Delta\xi,
\eeq 
where $d\N=4\pi r^2 c\nbg dt = 4\pi c \N d\ln r$ is the particle number (isotropic equivalent) passing through the wave during $dt$, and we used the fact that $\Ep^2r^2\propto r^{-4}$. \Eqs~(\ref{eq:gu2}) and (\ref{eq:cu})  give the same $\gu$. 

The consideration of the $C^+$ flow or energy conservation gives a simple solution for $\gamma(\xi)$ along the plateau. In particular, using $\c\ll 1$ and $\sin^2(\omega\xiin)\approx \Rm^4/r^4$, we find from \Eq~(\ref{eq:xi_kappa}),
\beq
\label{eq:c_profile}
  \gamma(\xi)\approx\frac{1}{2\c} 
  \approx \frac{c\sigbg}{r}(\xi-\xiin) \qquad (\xiin<\xi<\xish).
\eeq 
The solution reproduces $\gamma(\xi)$ found in the simulations (Figure~5). It ends at the shock with $\gamma(\xish)=\gu$.
 
Note that the plasma magnetization in the wave $\sigma=\c\sigbg$ is lowest just upstream of the shock, $\sigu=\cu\sigbg\approx\sigbg/2\gu$. Its minimum value is reached soon after the wave crosses $\Rm$: $\sigma_{\min}\sim\zeta\sigbg=\omega\Rm/c\gg 1$. This minimum value is independent of $\sigbg$.

\subsection{Beyond single-fluid MHD}
\label{MHD_validity}

The accelerated flow experiences dramatic expansion in the region $\xish<\xi<\xiin$, by the factor of $\c^{-1}\gg 1$, and the gyro-frequency in the fluid frame $\tomB=\c\omB$ drops. Note that the proper time $\tilde t$ (measured in fluid frame) slows down by a similar factor, $d\tilde t=dt/\gamma\approx 2\c\, dt$. 

Consider now the profile of $\gamma(\xi)\gg 1$ (\Eq~\ref{eq:c_profile}). A fluid element crosses $d\xi$ in time $dt=d\xi/(1-\beta)\approx d\xi/2$, and we find that $\gamma$ changes along the fluid streamline on the proper timescale $\tilde{t}_{\rm ev}$ given by
\beq
   \tilde{t}_{\rm ev}^{-1} \equiv \frac{d\ln\gamma}{d\tilde t} 
   =\frac{d\gamma}{dt} \approx  2\frac{d\gamma}{d\xi}
   \approx \frac{2c\sigbg}{r}.
\eeq
MHD description assumes that the gyration timescale $\tomB^{-1}\approx(\omB/2\gamma)^{-1}$ is much shorter than $\tilde{t}_{\rm ev}$. This condition may be stated as
\beq
   \tomB \tilde{t}_{\rm ev} \approx \frac{\omB r}{4\gamma\sigbg c} = \frac{\omegam r}{4\gamma\sigm c}\gg 1.
\eeq
It requires $\gamma\ll \gMHD$, where
\beq
\label{eq:gMHD}
   \gMHD\approx \frac{\omegam  r}{4 \sigm c}  = \frac{\pi e \N r}{\mu}.
\eeq
The MHD solution for the wave propagation problem remains mathematically well behaved even when $\gamma$ reaches arbitrary high values. However, its applicability to a real plasma is vindicated only if $\gamma\ll\gMHD$, which roughly corresponds to $\zeta\gg (2\omega/\omegam)^{1/2}$. Our sample numerical model with $\N=10^{39}$ had the peak $\gamma\approx 10^5$, just below $\gMHD\approx 2\times 10^5$, so the MHD description is marginally valid near the peak and accurate at other radii. The model with $\N=10^{37}$ developed $\gamma\gg\gMHD$, breaking the MHD condition.

In waves with $\zeta\simlt (2\omega/\omegam)^{1/2}$, where $\gamma$ reaches $\gMHD$,  the $e^+$ and $e^-$ motions are no longer the common $\bE\times\bB$ drift, and they are no longer coupled to the magnetic field via gyration. The momenta of unmagnetized $e^+$ and $e^-$ become significantly different and a two-fluid description is required to formulate the electric current $\bj$ and its effect on the electromagnetic field of the wave. Main features of the two-fluid solution may be anticipated using the following two considerations.

(1) The symmetry of $e^+$ and $e^-$ motions in an electromagnetic wave implies that they have equal velocities $\beta_z$ along the wavevector $\bk$ and opposite transverse velocities $\pm \beta_x$ with $\beta_y=0$ \citep{Beloborodov22}. We here use local Cartesian coordinates $x,y,z$ with the $z$ axis along $\be_r$, $y$ along $\be_\theta$, and $x$ along $-\be_\phi$. The symmetry implies electric current $\bj=j\,\be_x$ where $j=cen\beta_x$. Where vacuum propagation would give $E^2>B^2$, current is excited in the plasma to limit the growth of $E^2$ to $E^2\approx B^2$. The two-fluid plasma with $\gamma>\gMHD$ can enforce $E^2\simlt B^2$ similar to normal MHD. The ceiling of $E^2\approx B^2$ corresponds to the plateau $E\approx -\Bbg/2$, and $j$ needed to sustain $E\approx -\Bbg/2$ is found from \Eq~(\ref{eq:j_sw}):
\beq
\label{eq:j3}
     j\approx - \frac{\partial_t(rE)_\xi}{2\pi r} \approx \frac{c}{4\pi r} \frac{d(r\Bbg)}{dr}
       \approx -\frac{c\Bbg}{2\pi r}.
\eeq
The plasma accelerated in the wave flows with speed $\beta_z\approx -1$ and density $n\approx \nbg/(1-\beta_z)\approx \nbg/2$ (\Eq~\ref{eq:sw}), so the current $j$ corresponds to
\beq
\label{eq:bx}
    \beta_x = \frac{j}{ecn}\approx -\frac{\Bbg}{\pi er \nbg} =- \frac{\mu}{\pi e \N r}. 
\eeq
Note that $|\beta_x|\approx \gMHD^{-1}$. If $E^2$ significantly exceeded $B^2$, particles would be accelerated to a much greater $\beta_x$. This cannot happen, because the small $\beta_x\approx -\gMHD^{-1}$ already gives $j$ sufficient to enforce the ceiling $E^2\approx B^2$. Thus, the wave profile $E(\xi)$ in the two-fluid regime should evolve similarly to MHD --- an excess of $E^2>B^2$ will be shaved off at $E\approx -\Bbg/2$.

(2) Reducing $E^2$ to satisfy the ceiling of $E^2\approx B^2$ implies a well-defined loss of electromagnetic energy of the wave and the corresponding energy gain by the plasma (\Sect~\ref{gu}). Energy conservation then determines the profile of $\gamma$ ahead of the shock just like in the MHD regime, which gives \Eq~(\ref{eq:c_profile}) for $\gamma(\xi)$.

So, we expect that the analytical expressions for $\gamma$, $\beta_x$, and $u_x=\gamma\beta_x$ derived in MHD should carry over to the unmagnetized (two-fluid) regime.  The transition from MHD to the two-fluid regime occurs at $|u_x|\sim 1$. The MHD condition $|u_x|\ll 1$ may also be stated as 
\beq
  \gamma\ll \frac{1}{|\beta_x|}\approx  \frac{\pi e \N r}{\mu} 
  \quad ({\rm MHD}),
\eeq
which is equivalent to $\gamma\ll\gMHD$ (\Eq~\ref{eq:gMHD}). Note that $|u_x|\ll 1$ implies $\gamma\approx \gamma_z\equiv (1-\beta_z^2)^{-1/2}$, i.e. the fluid energy arises from its bulk motion along $z$ while the internal transverse motions of $e^\pm$ are negligible. By contrast, in the unmagnetized regime the $e^\pm$ develop $|u_x|\gg 1$, i.e. the internal motions become relativistic and significantly contribute to the plasma energy. This change will affect the shock jump conditions; however the shock speed should anyway be regulated to the value given in \Eq~(\ref{eq:shock_speed1})  below.

It is also instructive to look at the dynamical equations describing the two-fluid $e^\pm$ motions. Taking into account the symmetry of $e^\pm$ velocities, it is sufficient to consider the positrons. The energy equation in the plateau region $E\approx -\Bbg/2$ reads
\beq
\label{eq:gt}
    \frac{d\gamma}{dt} = \frac{eE \beta_x}{\me c}  \approx -\frac{\omB}{2}\,\beta_x,
\eeq 
where $\omB=e\Bbg/\me c$ and $d/dt$ is taken along the particle trajectory, so $dt=d\xi/(1-\beta_z)\approx d\xi/2$. Substituting $\beta_x$ from \Eq~(\ref{eq:bx}), one finds 
\beq
   \frac{d\gamma}{d\xi}=\frac{c\sigbg}{r},
\eeq
recovering the solution~(\ref{eq:c_profile}) for $\gamma(\xi)$.

Sustaining $\beta_x$ according to \Eq~(\ref{eq:bx}) requires a small mismatch between $\beta_z$ and $\bD=E/B$, as seen from the $x$-momentum equation,
\beq
    \frac{du_x}{dt} = \frac{e}{\me c}\left(E-\beta_z B\right) 
     \approx \frac{\omB}{2}(\bD-\beta_z).
\eeq 
Let us define $\Delta\equiv -(B+E)/B$ as a measure for the deviation of $E$ from $-B$; $\Delta>0$ corresponds to $E^2>B^2$. It is easy to show that $\Delta>1+\beta_z\approx (2\gamma_z^2)^{-1}$ is required to achieve $u_x$ of the unmagnetized regime, so $du_x/d t \approx -(\omB/2)\Delta$. The electric current $j$ and the corresponding $\beta_x$ (\Eq~\ref{eq:bx}) will be sustained if $du_x/dt = \beta_x\, d\gamma/dt = -\beta_x^2\,\omB/2$, which requires 
 \beq
    \Delta \approx \beta_x^2\approx \left(\frac{\mu}{\pi e \N r}\right)^2\approx \frac{1}{2\gMHD^2}.
 \eeq
The tiny positive $\Delta$ shows that $E^2$ slightly exceeds $B^2$ in the unmagnetized regime. By contrast, in the MHD regime, $\Delta \approx -(2\gamma^2)^{-1}<0$, i.e. $E^2<B^2$. In both cases, the ceiling of $E^2\approx B^2$ is strongly enforced.

\subsection{Shock strength}
\label{strength}

The upstream flow is decelerated in the shock: we observed in the simulation a huge jump of the plasma proper density $q=\trhod/\trhou$, which implies a huge jump of Lorentz factor $\gu/\gd\gg 1$. By contrast, the electric field has a small jump during the peak of the monster shock: $\Ed-\Eu\approx \Ed+\Bbg/2$ is a small fraction of $\Eu\approx-\Bbg/2$. This means that the plasma immediately behind the shock still moves with a large speed: $\betad=\Ed/(\Bbg+\Ed)\approx -1$ and $\gd\gg1$. The plasma decelerates to $\gamma<\gd$ further  downstream where $E>\Ed$ significantly deviates from $-\Bbg/2$.

The ratio $\gu/\gd$ is regulated by the shock jump conditions. They require fast motion of the shock relative to the downstream plasma: $\gamma_{\rm sh|d}\approx \chi^{1/7}\sqrt{\sigu} \gg \sqrt{\sigu}$. Using $\gamma_{\rm sh|d}\approx 2\gsh\gd$ (note that $\betad\approx -1$ is opposite to $\bsh\approx 1$) one can show that $\gd\gg 1$, which corresponds to $\cd\ll 1$ and $\Ed+\Bbg/2\ll \Bbg/2$. 

The value of $\cd\ll 1$ may be derived as follows. The condition that $\Ed=\Em\sin(\omega\xish)$ stays close to $-\Bbg/2$  implies
\beq
\label{eq:xish}
   \sin(\omega\xish)\approx -\frac{\Bbg}{2\Em} 
    =-\frac{\Rm^2}{r^2}.
\eeq
This relation controls the shock position $\xish$ as a function of time $t\approx r/c$ and thus determines its speed $d\xish/dt=1-\bsh\approx (2\gsh^2)^{-1}$ (when the shock Lorentz factor $\gsh\gg 1$). This gives 
\beq 
\label{eq:shock_speed1}
   \gsh^2\approx \left(2\frac{d\xish}{dt}\right)^{-1}
   \approx \frac{\omega \Rm }{4c}\, x\, \sqrt{x^4-1}.
\eeq
On the other hand, the shock jump conditions require $d\xish/dt$ given in \Eq~(\ref{eq:shock_speed}). Substituting it into \Eq~(\ref{eq:shock_speed1}), we find
\beq
\label{eq:cd}
   \cd^7 \approx \frac{\zeta^2}{32\eta}\,x^{11}(x^4-1),
\eeq
where $\eta=(\sT\Bm/8\pi e)$.

The obtained $\cu$ and $\cd$ (\Eqs~(\ref{eq:cu}) and (\ref{eq:cd})) determine $q=\trhod/\trhou=\cd/\cu$:
\beq
\label{eq:q}
   q\approx \left[\frac{4(1-x^{-4})}{\zeta^5\, \eta\, x^{13}}\right]^{1/7} 
   \left(\pi - 2\arcsin\frac{1}{x^2}\right).
\eeq
This result is in excellent agreement with the numerical simulation (Figure~\ref{fig:kHz_q}). It loses accuracy at large radii where $\Ed$ significantly deviates from $-\Bbg/2$, so that the approximation~(\ref{eq:xish}) becomes invalid. This occurs when $\cd$ (\Eq~\ref{eq:cd}) increases to $\sim 1$, i.e. at radius
\beq
   x_1\approx \left(\frac{32\eta}{\zeta^2}\right)^{1/15}.
\eeq
Note also that $\cd\approx 1$ corresponds to $q\approx 2\gu$. For the wave simulated in \Sect~\ref{numerical} with $\N=10^{39}$ we find $x_1\approx 1.85$ which corresponds to $r_1=x_1\Rm\approx 2.6\times 10^8\,$cm.

The shock radiative parameter $\chi$ is given by \Eq~(\ref{eq:jump2}), as long as $\chi\gg 1$. Substitution of \Eqs~(\ref{eq:cu}) and (\ref{eq:cd}) for $\cu$ and $\cd$ yields
\beq
\label{eq:chi2}
   \chi=2^{3/2}\,\frac{\eta}{\zeta^2}\,\frac{(1-x^{-4})^{3/4}}{x^8}\left(\pi - 2\arcsin\frac{1}{x^2}\right)^{7/2}.
\eeq
It reproduces the value of $\chi$ observed in the simulation.

\subsection{Precursor effect}
\label{precursor}

In an accompanying paper we describe an additional effect: the shock emits  precursor waves into the upstream, which can significantly decelerate the upstream flow before it reaches the collisionless shock.

The precursor emission begins immediately after shock formation at time $t_\times\approx\Rm/c$ near the coordinate $\xitrough= 3\pi/2\omega$. As the wave expands to $r>\Rm$, the shock moves to $\xish>\xitrough$, so the precursor occupies the growing region 
\beq
    \frac{3\pi}{2\omega}=\xitrough<\xi<\xish. 
\eeq
Note that the plasma flow in the MHD wave crosses half of the plateau $E\approx -\Bbg/2$ ahead of the precursor, at $\xi<\xitrough$, and begins to interact with the precursor at $\xitrough$ with the Lorentz factor determined by \Eq~(\ref{eq:c_profile}),
\beq
\label{eq:gu_prec}
   \gamma(\xitrough,r) \approx \frac{\Rm^4}{\zeta\, r^4}\left(\frac{\pi}{2}-\arcsin\frac{\Rm^2}{r^2}\right).
\eeq
It is lower by a factor of 2 compared to $\gu$ defined previously at $\xish$ (\Eq~\ref{eq:cu}). In the accompanying paper we find that the precursor can decelerate the upstream flow as it moves from $\xitrough$ to $\xish$, so the flow comes to the shock with $\gamma(\xish)\ll\gamma(\xitrough)$ (and the shock radiative parameter $\chi$ may drop below unity). This pre-deceleration effect depends on somewhat uncertain efficiency of precursor emission, which has not been studied in the extreme regime of the monster shocks.

Regardless of the shock structure details, with or without the precursor, the basic picture remains the same: the upstream first develops the huge Lorentz factor $\gamma(\xitrough)$ [or $2\gamma(\xitrough)$ without the precursor] and then promptly radiates its energy, before completing the shock transition. The emitted radiation is in the X/gamma-ray band, as shown in \Sect~\ref{emission} below.

\subsection{Forward shock}
\label{FS}

The plateau evolution described above leads to erasing half of the wave oscillation where $E<0$. It does not affect the leading half $0<\omega\xi<\pi$ where $E>0$ (Figure~5). This leading part of the wave propagates with negligible distortions at radii $r\sim\Rm$, nearly as in vacuum. However, at larger $r\gg\Rm$ the profile of $E(\xi)>0$ gradually steepens at the leading edge $\xi=0$ and eventually forms a forward shock. Below we find radius $\RF$ where this forward shock is launched. 

The part of the wave with $E>0$ has $\c>1$ and satisfies $\c^3\sigbg\gg 1$. This implies small bending of characteristics, $d\xi_+/dt\ll 1$, and the evolution of $\c$ along $C^+$ follows the simple relation $\c^2=1+2\K r^2$ (Appendix~\ref{cold_wave}).

In the zero order of the small parameter  $d\xi_+/dt$, $C^+$ are described by $\xi_+=\xiin$ or $r_+=c(t-\xiin)$. The correction $\xi_+-\xiin=(\c-1)^2/8c\D\K^2\c$ (\Eq~\ref{eq:flow_xi+}) in the leading order may be found iteratively by substituting $\c^2=1+2\K r^2$ with the zero-order $r=c(t-\xiin)$. This gives
\beq
\label{eq:small_bending}
   \c^2 = 1+2Kc^2(t-\xiin)^2, \quad
   \c \left.\frac{\partial\c}{\partial\xiin}\right|_t = \frac{dK}{d\xiin} r^2 - 2\K r c.
\eeq
One can now examine the flow of $C^+$ characteristics $\xiin\rightarrow\xi_+(\xiin,t)$ and identify where $(\partial\xi_+/\partial\xiin)_t$ vanishes, launching a shock. For the $C^+$ flow with $\sigbg\c^3\gg 1$ one can use \Eq~(\ref{eq:small_bending}) and find
\beq
\label{eq:dxi+_dxiin_}
  \left.\frac{\partial\xi_+}{\partial\xiin}\right|_t = 1-\frac{r^3}{2D\c^3} \left[ 
  \frac{dK}{d\xiin} \frac{r^3}{c} \frac{(3\c+1)}{(\c+1)^3}+1\right].
\eeq

We here focus on the interval $0<\omega\xiin<\pi$ and observe that $\partial\xi_+/\partial\xiin$ drops fastest at the leading edge $\xiin=0$, where $\c=1$ is minimum and $dK/d\xiin=\Km\omega=\omega/2\Rm^2$ is maximum. We find at the leading edge (using $r^3d\K/d\xiin =\omega r^3/2\Rm^2\gg c$),
\beq
    \left.\frac{\partial\xi_+}{\partial\xiin}\right|_t=1-\frac{\omega r^6}{8c\D\Rm^2}
    =1-\frac{\zeta}{8}\,\frac{r^6}{\Rm^6}.
\eeq
Hence, the forward shock forms at radius
\beq
\label{eq:RF}
   \RF=\left(\frac{8}{\zeta}\right)^{1/6} \Rm 
   \approx  3 \times 10^9 \frac{\mu_{33}^{1/2} } { \N_{37}^{1/6} \nu_4^{1/6} L_{43}^{1/12} } {\rm ~cm}.
\eeq
The wave now carries two shocks, at $\xi\approx 0$ and $\xi>2\pi/\omega$, which  slowly shift to larger $\xi$. The nascent forward shock at $r=\RF$ is weak. It becomes ultrarelativistic at $r\gg\RF$ and later turns into a strong blast wave expanding into the wind outside the magnetosphere.


\section{Gamma-rays from monster shocks}
\label{emission}

First, consider shocks with neglected effects of the precursor on the upstream flow.
As the plasma flow crosses the shock, it radiates its energy in curvature photons. The spectrum of curvature radiation from a particle with Lorentz factor $\gamma$ cuts off  exponentially at $\omega>\omc$ \citep{Landau75}, where 
\beq
\label{eq:omc}
  \omc=\left(\frac{3}{2}\right)^{3/2} \left(\frac{c\,\dEe}{e^2}\right)^{1/2} \gamma.
\eeq
The radiated power $\dEe=\dot{\gamma}mc^2$ is Lorentz invariant, and may be evaluated in the shock frame, $\dot{\gamma}=|d\gamma'/dt'|$, using the solution derived in Appendix~\ref{shock_structure}. It shows that $\gamma'$ drops from the upstream value $\gu'\approx 2\gu\gsh\approx \gsh/\cu$ with rate 
\beq
  \dot{\gamma}
  \approx \frac{4}{3}\,\frac{\gu'}{\sqrt{\sigu}} \frac{e\tB_{\rm u}}{mc}\, (\sqrt{g}-g^2)
  \approx \frac{4}{3}\,\frac{\gsh  \omB}{\sqrt{\sigu}} \, (\sqrt{g}-g^2),
\eeq
where $g=\gamma'/\gu'$, $\omB=e\Bbg/mc$, and $\tB_{\rm u}=\cu\Bbg$. The emission frequency $\omc$ is maximum at $g=(5/8)^{2/3}$,
\beq
\label{eq:omc1}
   \omc^{\max}\approx 0.88\,\gu \left(\frac{c\,\omB}{r_e} \frac{\gsh}{\sqrt{\sigu}} \right)^{1/2}.
\eeq
Substituting the solutions for $\gu(r)$ (\Eq~\ref{eq:cu}), $\gsh(r)$ (\Eq~\ref{eq:shock_speed1}), and $\sigu=\cu\sigbg\approx\sigbg/2\gu$, we find
\beq
\label{eq:nuc}
    \epc\equiv\frac{\hbar\omc^{\max}}{mc^2}\approx \frac{1}{\zeta}
    \left(\frac{\Bm}{\alpha_f B_Q}\right)^{1/2} f_c(x),
\eeq
where $\alpha_f=e^2/\hbar c\approx 1/137$, $B_Q=m^2c^3/\hbar e\approx 4.4\times 10^{13}\,$G, $\Bm=\Bbg(\Rm)$, $x=r/\Rm$, and 
\beq
   f_c(x)=\frac{0.88(\pi-2\arcsin x^{-2})^{5/4}(1-x^{-4})^{1/8}}{2^{1/4}x^5}.
\eeq
One can see that $\omc^{\max}$ is in the far gamma-ray band. 

Next, consider emission from shocks affected by the precursor (\Sect~\ref{precursor}). Main emission now occurs where the upstream flow enters the precursor and begins deceleration with $\gamma\approx\gamma(\xitrough)$ given by \Eq~(\ref{eq:gu_prec}). The emission frequency is still given by \Eq~(\ref{eq:omc}), but now with $\dEe$ being the emitted power that results from the particle interaction with the precursor. Using $d\xi=(1-\beta)dt\approx 2dt$ along the plasma streamline, one can see that $\delta\xidec\equiv 2\gamma\me c^2/\dEe$ defines a characteristic deceleration scale, and we rewrite \Eq~(\ref{eq:omc}) as
\beq
\label{eq:omc_prec}
  \omega_c = \frac{3^{3/2}}{2}\,\gamma^{3/2}\left(\frac{c}{r_e\,\delta\xidec}\right)^{1/2}.
\eeq
During the main phase of shock evolution, the upstream flow reaches the peak Lorentz factor $\gamma(\xitrough)\approx (2\zeta)^{-1}$ and radiates photons with dimensionless energy
\beq
  \epc\approx \frac{ 8\times 10^{-9}\,\nu_4^{1/2} }{ (\omega\,\delta\xidec)^{1/2} \zeta^{3/2}},
\eeq 
where we have normalized $\delta\xidec$ to the precursor width $\xish-\xitrough\sim \omega^{-1}$. Note that the precursor deceleration effect is strong if $\delta\xidec\ll \omega^{-1}$, and so $\epc\gg 10^{-8}\zeta^{-3/2}$. This gives the characteristic $\epc$ in the gamma-ray band.


\section{Discussion}
\label{discussion}

\subsection{Formation of shocks}

The magnetospheres of neutron stars have a huge magnetization parameter $\sigbg$, and therefore their low-frequency perturbations are often described as FFE waves, which propagate with the speed of light. This description is excellent near the star, however it fails when perturbations propagate to larger radii and grow in relative amplitude $E/\Bbg$. Then, FFE  becomes remarkably self-destructive: it pushes itself out of the realm of its applicability $E^2<B^2$, and the wave dynamics in the FFE limit $\sigbg\rightarrow\infty$ becomes undefined. Therefore, waves should be described using the full MHD framework, with an arbitrarily large but finite $\sigbg$.

In particular, compressive MHD waves (vacuum radio waves in the FFE limit) steepen into shocks at radius $\Rm$ (\Eq~\ref{eq:Rm}) for any high $\sigbg$. The higher $\sigbg$ the stronger the shock at $\Rm$, as we have demonstrated by solving the MHD wave equation. As an example, we presented the evolution of a 10-kHz wave with an initial sine profile $E=\Em\sin(\omega\xi)$, where $\xi=t-r/c$. We have also derived the wave evolution analytically to show how it depends on the wave frequency $\omega$ and power $L$. Our main conclusions are as follows.\footnote{These conclusions hold for kHz perturbations typically excited in magnetars, neutron star binaries, and neutron star collapse. The results are different for MHz-GHz waves \citep{Beloborodov23}.}

1. When the wave reaches $\Rm\sim 10^8$\,cm it suddenly begins to pull the background magnetosphere toward the star, creating an ultrarelativistic flow at the wave oscillation phase near $3\pi/2$ (trough), where $E^2$ approaches $B^2$ (Figure~5). The flow acceleration at the trough of the wave is accompanied by plasma expansion, reducing the local magnetization parameter $\sigma$, so the wave propagation slows down at the trough and the wave ``stumbles,'' steepening into a shock. The accelerated plasma flow forms the upstream of the shock. It develops a huge Lorentz factor $\gu\sim c\sigm/\omega\Rm$, where $\sigm\equiv\sigbg(\Rm)$. The peak $\gu$ is reached when the wave crosses $r\approx 1.15\Rm$, and here the local magnetization parameter of the flow is reduced  to $\sigu\sim \sigm/\gu\sim \omega\Rm/c$, which independent of $\sigbg$.

2. The monster shock has an unusual micro-structure. The upstream flow experiences strong radiative losses before completing a single Larmor rotation, i.e. before forming the downstream MHD flow. This affects the shock jump conditions. In particular, we found that the shock Lorentz factor relative to the downstream plasma is $\gamma_{\rm sh|d}\approx(1+\chi)^{1/7}\sqrt{\sigu}$, where $\chi$  is a new dimensionless radiative parameter (\Eq~\ref{eq:chi1}). The shock structure is further complicated by the precursor emission, which can induce radiative losses of the upstream flow ahead of the shock; this effect will be described elsewhere.

3. As the kHz wave propagates beyond $\Rm$, half of its oscillation ($\pi<\omega\xi<2\pi$) becomes erased (Figure~5). Thus the wave loses half of its energy after crossing $\Rm$. The large wave energy per magnetospheric particle corresponds to the huge Lorentz factor gained by the plasma. The plasma radiates the gained energy in the ultra-relativistic shock, leading to a bright X-ray burst.

4. Waves accelerating the plasma to $\gamma\sim(2\omB/\omega)^{1/2}$ enter the two-fluid regime with unmagnetized particles. We argued in \Sect~\ref{MHD_validity} that the main MHD results will carry over to the two-fluid regime. In particular, the Lorentz factor of upstream particles will be given by the same expression $\gamma\sim c\sigm/\omega\Rm$, reaching enormous values in powerful low-frequency waves.

5. The leading half of the wave oscillation ($\omega\xi<\pi$) never approaches the condition $E^2=B^2$. It crosses $\Rm$ without any significant changes. A wave consisting of only this half-oscillation would not develop the monster radiative shock at $\Rm$. However, in any case, at a larger radius $\RF$ (a few times $10^9\,$cm, see \Eq~(\ref{eq:RF})), the leading edge of the wave steepens into a forward shock.

Both $\Rm$ and $\RF$ are well inside the typical light cylinder of a magnetar, $\RLC\sim 10^{10}\,$cm. Thus, strong kHz waves from magnetars inevitably launch shocks inside the magnetosphere. The monster shock forms at $\omega\xi\approx 3\pi/2$; then it weakens and shifts to $\omega\xi\approx 2\pi$. The forward shock forms at $\xi\approx 0$ and gets stronger at large radii; it will evolve into an ultrarelativistic blast wave expanding into the magnetar wind
(this evolution will be further described elsewhere).

Our results clarify the relation between two pictures proposed for electromagnetic ejecta from magnetars: a vacuum-like electromagnetic wave \citep{Lyubarsky14} vs. a blast wave in the wind \citep{Beloborodov17b,Beloborodov20}. Blast wave formation is important for the theory of fast radio bursts (FRBs, see \citealt{Lyubarsky21} for a review), because it produces semi-coherent radio emission with sub-ms duration. Kinetic simulations demonstrating the efficiency and polarization of this emission are found in \cite{Sironi21}. In addition, \cite{Thompson23} recently proposed that the blast wave may produce radio waves by a different mechanism if it expands into a turbulent medium.

\subsection{Compton drag}
\label{drag}

Our calculations of magnetospheric MHD waves neglected Compton drag exerted by ambient radiation. Radiation flows from magnetars with luminosities $L_\star\sim 10^{35}\,$erg/s, peaking at photon energies $\eph mc^2\sim 1\,$keV \citep{Kaspi17}. When plasma in the wave develops Lorentz factor $\gamma\gg 1$, it upscatters the keV photons to energies $\esc\sim\gamma^2\eph$ as long as $\gamma<\eph^{-1}$. This gives a maximum power of scattered radiation $\Lsc^{\max}\sim \gamma^2L_\star\sim \eph^{-2}L_\star\sim 3\times 10^{40}\,$erg/s (this upper bound assumes that each keV photon is upscattered). Waves with power $L\gg \Lsc$ are weakly affected by Compton drag. When $\gamma>\eph^{-1}$ scattering occurs with a smaller cross section $\sigsc\sim \sT/\gamma\eph$ and $\esc\sim \gamma$. Using the scattering optical depth $\tausc\sim \sigsc\N/\Rm^2$, we find 
\beq
  \frac{L_{\rm sc}}{L} \sim \frac{L_\star \varepsilon_{\rm sc} \tausc }{L\, \varepsilon_\star}
   \sim \frac{\sqrt{8} \sT \N L_\star}{\mu (cL)^{1/2}\eph^2}
   \sim 10^{-4}\,\frac{L_{\star,35}}{\varepsilon_{\star,-3}^2}\,\frac{\N_{39}}{\mu_{33}L_{44}^{1/2}}.
\eeq
This again gives $L_{\rm sc}\ll L$, even if the density parameter $\N=n r^3$ is increased by secondary $e^\pm$ creation (which mainly occurs behind the wave, not inside it, see \Sect~\ref{pairs} below). Thus, we conclude that the magnetar radiation does not prevent the enormous plasma acceleration in magnetosonic waves at $r\approx \Rm$.

\subsection{Wave evolution at $\theta\neq\pi/2$}
\label{oblique}

\begin{figure}[t]
\includegraphics[width=0.47\textwidth]{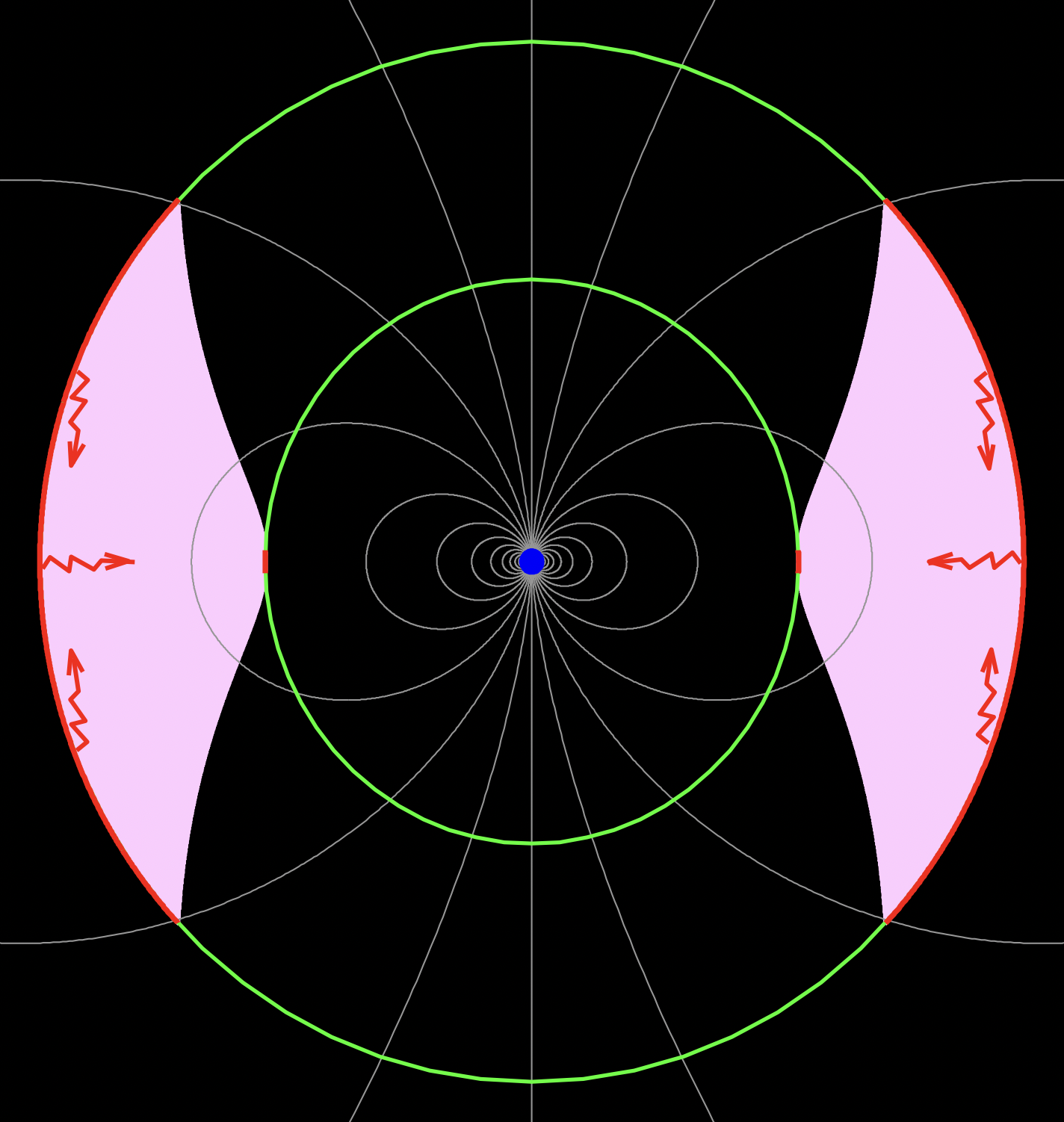} 
\caption{Expansion of a spherical wave front through the magnetosphere. The front thickness (wave duration) is much smaller than $r/c$. Its shock-free part ($\sin\theta<\sin\thsh$) is plotted in green and the part carrying the shock is in red. The region swept by the shock is shaded in pink; its boundary $\theta(r)$ is defined by \Eq~(\ref{eq:rsh}). The wave front is shown at two moments: (1) when it crosses radius $\Rm$ (the shock has formed in the equatorial plane, $\theta=\pi/2$); (2) when it expands to a larger radius (the shock now occupies a range of $\theta$). The shock radiates gamma-rays (red arrows) beamed along $\bb_{\rm D}$ (\Eq~\ref{eq:bD});
most of them convert to $e^\pm$ pairs.}
\label{fig:magnetosphere}
 \end{figure}

In this paper, we calculated the wave propagation in the equatorial plane of a dipole magnetosphere, $\theta=\pi/2$. The obtained solution shows a clear physical picture of the wave evolution. It includes two facts that help extend the picture to $\theta\neq \pi/2$:

(1) The MHD wave is well described as a vacuum electromagnetic wave superimposed on the background dipole magnetosphere until this superposition hits the condition $E^2=B^2$, launching the monster shock. In the equatorial plane, this condition is reached at $\Rm$. In other parts of the magnetosphere,  one can find the location of shock formation from the same condition $E^2\approx B^2$, which corresponds to drift speed $|\bb_{\rm D}|\rightarrow 1$. The drift velocity is given by
\beq
   \bb_{\rm D}=\frac{\bE\times\bB}{B^2} = \frac{E\left(B_{\rm bg,\theta}+E\right)\be_r 
    - EB_{{\rm bg},r}\be_\theta}{\Bbg^2+2EB_{\rm bg,\theta}+E^2},
\eeq
with $\bE=-E\be_\phi$ and $\bB=\bBbg+E\be_\theta$. The wave propagates as in vacuum until $\bD^2=E^2/B^2$ approaches unity, which corresponds to  $\Bbg^2+2EB_{\rm bg,\theta}=0$. This condition is first approached at the trough of the wave, $E=-\Em$. It defines the critical surface $\rsh(\theta)$: 
\beq
\label{eq:rsh}
   \Em=\frac{\Bbg^2}{2 B_{\rm bg,\theta}} \quad \Rightarrow \;\;
   \rsh(\theta)=\Rm\sqrt{\frac{4-3\sin^2\theta}{\sin\theta}}.
\eeq
A spherical wave with approximately isotropic power reaches this surface first at $\theta=\pi/2$, $r=\Rm$. As the wave continues its propagation to $r>\Rm$, the monster shock develops in a range of latitudes, $\sin\thsh<\sin\theta\leq1$ (Figure~\ref{fig:magnetosphere}), where $\sin\thsh(r)$ is found by solving \Eq~(\ref{eq:rsh}) for $\sin\theta$. 

The condition $E^2\rightarrow B^2$ corresponds to the plasma velocity $\bb$ approaching the unit vector,\footnote{The oblique (non-radial) plasma motion in the wave at $\theta\neq \pi/2$ develops a velocity component $\bb_\parallel$ parallel to $\bB$, as discussed in the accompanying paper \citep{Beloborodov23}. Then, the accurate total four-velocity is $\bu=\bu_\parallel+\bu_{\rm D}$. However, $u_\parallel\ll u_{\rm D}$ when $E^2\rightarrow B^2$.}  
\begin{eqnarray}
\nonumber
   \bb_{\rm D} &=& \frac{(B_{{\rm bg},r}^2-B_{\rm bg,\theta}^2)\be_r+2B_{{\rm bg},r}B_{\rm bg,\theta}\be_\theta}{\Bbg^2}  
   \quad (E^2=B^2)
   \\
   & = & \frac{(4-5\sin^2\theta)\be_r+2\sin(2\theta)\be_\theta}{4-3\sin^2\theta}.
 \label{eq:bD}
\end{eqnarray}
This unit vector represents the direction of the accelerated plasma flow. Note that the radial component of $\bb_{\rm D}$ changes sign at $\sin\theta=2/\sqrt{5}$.
    
(2) The obtained solution at $\theta=\pi/2$ demonstrates that the wave evolution outside $\Rm$ follows a simple principle: the part $\Delta\xi$ of the wave oscillation that has hit the ceiling $E^2=B^2$ forms the plateau of $E\approx \Ep$ while the rest of the oscillation profile $\bE(\xi)$ follows nearly vacuum propagation (until the wave approaches $\RF\gg\Rm$). A similar description can be used at $\theta\neq\pi/2$. 
The plateau electric field is set by the condition $E^2\approx B^2$: 
\beq
  \bE_{\rm p}(r,\theta) = \frac{\Bbg^2}{2B_{\rm bg,\theta}} \,\be_\phi 
  =\frac{\mu}{2r^3}\, \frac{\rsh^2}{\Rm^2} \,\be_\phi.
\eeq
It is sustained by electric current $\bj_{\rm p} = -\partial_r(r\bE_{\rm p})/2\pi r$. The plasma passing through the wave is accelerated on the plateau with rate $d\gamma/dt=\bE_{\rm p}\cdot\bj_{\rm p}/\rho c^2$, which gives
\beq
  \frac{d\gamma}{d\xi} = \frac{\bE_{\rm p}\cdot \bj_{\rm p}}{\rhobg c^2} 
  = -\frac{\partial_r(r^2\Ep^2)}{4\pi c \rhobg  r^2} 
  = \frac{c\sigbg}{r}\,\frac{\Bbg^2}{B_{\rm bg,\theta}^2},
\eeq
where we used $d\xi=(1-\beta_r)dt=(\rhobg/\rho)dt$ along the plasma streamline. A plateau of width $W_{\rm p}=c\Delta\xi$
gives
\begin{equation}
  \gamma \approx \sigbg \frac{W_{\rm p}}{r} \frac{\Bbg^2}{B_{\rm bg,\theta}^2}.
\end{equation}
The acceleration region $W_{\rm p}$ with $\bE\approx \bE_{\rm p}$ and current $\bj_{\rm p}$ appears at $r=\rsh$ and grows at $r>\rsh$. For example, waves with initial profile $E(\xi)=\Em\sin(\omega\xi)$  develop plateaus with $W_{\rm p}(r)\approx (c/\omega)[\pi -2 \arcsin(\rsh^2/r^2)]$.

As a result of the wave evolution at $r>r_\times$, the parts of the wave profile with $\bE\cdot (\bk\times\bBbg)<0$ become erased and replaced by the plateau with $\Ep/\Em\rightarrow 0$ at $r\gg\rsh$.  In axisymmetric waves $\bE$ oscillates along $\be_{\phi}$, and so the parts with $E_\phi>0$ become erased. In particular, oscillating kHz waves with average $\overline{E}_\phi\approx 0$ lose half of their energy. The lost energy is radiated by the shock.

\subsection{Pair creation and emission of X-ray bursts}
\label{pairs}

The monster shock radiates high-energy photons (\Sect~\ref{emission}), which can convert to $e^\pm$ pairs. The upstream flow entering the equatorial shock moves along $-\be_r$ with a high Lorentz factor $\gamma$, so its gamma-ray emission is beamed toward the star. These gamma-rays propagate perpendicular to $\bBbg$ and will convert to $e^\pm$ pairs. The mean free path for conversion is given by \citep{Erber66},
\beq
   \lph(r)\approx \frac{4.4\,r_e}{\alpha_f^2} \frac{B_Q}{\Bbg}\exp\left(
   \frac{8}{3\epc}\frac{B_Q}{\Bbg} \right).
\eeq
For our sample model in \Sect~\ref{numerical}, $\lph(\Rm)\gg \Rm$ and $\lph(r)$ will be reduced below $r$ when the gamma-rays propagate to $r\ll\Rm$ where $\Bbg\gg\Bm$. 

Magnetosonic waves accelerate the plasma radially only in the equatorial plane of the magnetosphere. Waves at different latitudes accelerate the plasma obliquely (\Sect~\ref{oblique}), so the gamma-ray emission is oblique, not beamed toward the star, and can avoid absorption by the ultrastrong field. Instead, the oblique gamma-rays will convert to $e^\pm$ pairs via photon-photon collisions, as they propagate toward the equatorial plane and collide with the symmetric gamma-rays from the lower hemisphere. The collisions will be efficient because of the broad spectrum of curvature radiation (its half-width extends from $0.01\epc$ to $1.5\epc$, see e.g. \cite{Longair94}), so the gamma-rays will find counterparts near the threshold for $e^\pm$ creation with a large cross section $\sim 0.1\sT$.  This leads to a high pair production rate, and the region behind the shock will become populated with optically thick $e^\pm$ plasma.

The created $e^\pm$ plasma will cool by emitting softer synchrotron photons and by scattering the photons as they diffuse out of the $e^\pm$ cloud behind the shock. Radiation production is controlled by the compactness parameter, 
\beq
   \lm=\frac{\sT L}{8\pi \me c^3\Rm }\approx 7.7\times 10^4\,L_{43}^{5/4}\mu_{33}^{-1/2}.
\eeq
The high compactness implies a large optical depth to photon-photon collisions, so most gamma-rays should convert to pairs, which experience fast radiative cooling. Thus, the shock emission will be reprocessed to photons of lower energies. The luminosity of the resulting X-ray burst is comparable to the power of the original kHz wave. The burst resembles radiative processes in compact magnetic flares simulated in \cite{Beloborodov21a}; its spectrum can be found with similar detailed radiative transfer simulations, which we leave for future work. 

A minimum duration of the burst is set by the light crossing time $r/c$, where $r$ is a few $\Rm$. A typical value of this minimum duration is $\sim 10\,$ms. Bursts with high $\lm$ may last longer, as it takes time for the reprocessed X-rays to diffuse out of the optically thick $e^\pm$ cloud. In addition, starquakes may produce multiple shocks, creating multiple bursts. Such composite bursts can last $\sim 100$\,ms, depending on the crustal quake coupling to the neutron star core \citep{Bransgrove20}.

\subsection{X-ray precursor of a binary merger}
\label{binary}

Consider a neutron-star binary with masses $M_1$ and $M_2$ separated by distance $2a$. The reduced mass of the binary is $M=M_1M_2/(M_1+M_2)$, and its orbital angular velocity is
\beq
   \Omega\approx \frac{c}{r_g}\left(\frac{a}{r_g}\right)^{-3/2}, \qquad r_g\equiv \frac{GM}{c^2}. 
\eeq
In a tight binary (nearing merger) $\Omega$ is in the kHz band.

Suppose both stars are strongly magnetized. An interface between their magnetospheres forms where their pressures balance. For instance, consider two stars with equal magnetic dipole moments $|\boldsymbol{\mu}_1|=|\boldsymbol{\mu}_2|=\mu$. Then, the interface is in the middle, at the distance $a$ from each star, and the magnetic field at the interface is $B_i\approx \mu/a^3$.

If the orbiting stars are not in synchronous rotation, there is differential rotation between the two magnetospheres with an angular frequency $\Omega_{\rm diff}$ comparable to the orbital $\Omega$. The two magnetic moments $\boldsymbol{\mu}_1$ and $\boldsymbol{\mu}_2$ in general are not aligned, and then the magnetic pressure at the interface will oscillate with frequency $\Omega_{\rm diff}$. The pressure variation timescale $\sim\Omega_{\rm diff}^{-1}$ exceeds the Alfv\'en crossing time (a few $a/c$), which reduces the efficiency of low-frequency wave excitation. However, the interface is also a source of higher-frequency waves, because it is prone to instabilities of the Kelvin-Helmholtz/diocotron type. The instability will generate vortices of sizes up to a fraction of $a$ (with frequencies $\omega>c/a$), creating traction between the rotating magnetospheres.

The turbulent region around the interface will emit both Alfv\'en and magnetosonic waves, and further investigation of this process requires numerical simulations. An upper bound on the power deposited into the turbulence may be roughly estimated as $L_{\rm max} \sim \Omega_{\rm diff} a^3 B_i^2/8\pi$. The power of magnetosonic wave emission $L\sim ca^2(\delta B)^2/8\pi < L_{\max}$ implies an upper limit on the emission amplitude,
\beq 
  b\equiv \frac{\delta B}{B}<\left(\frac{\Omega_{\rm diff} R}{c}\right)^{1/2}
  \sim \left(\frac{r_g}{a}\right)^{1/4}\left(\frac{\Omega_{\rm diff}}{\Omega}\right)^{1/2}.
\eeq

In a more general case the magnetic dipole moments of the two stars are not equal, $\mu_1<\mu_2$. Then the interface will form at distance $R$ from $M_1$, defined by
\beq
  \frac{\mu_1}{R^3}\approx\frac{\mu_2}{(2a-R)^3}.
\eeq
The wave source will have the size of $\sim R<a$, and its power may be estimated as
\beq
   L\sim c R^2\,\frac{b^2B_i^2}{8\pi}.
\eeq
The emitted magnetosonic waves will steepen into shocks at radius $\Rm\sim (2b)^{-1/2}R$ (if it is inside the light cylinder of the binary, $\Rm<c/\Omega$).
  
The emission of kHz waves accompanied by shocks and turbulent dissipation will continue as long as the two stars orbit each other, until they eventually merge. The binary loses orbital energy to gravitational wave emission and shrinks on the timescale 
\beq
   t\sim \frac{r_g}{c}\left(\frac{a}{r_g}\right)^4,
\eeq
which is also comparable to the time remaining until the merger. The wave energy generated during the remaining time $t$ may be estimated as $tL$, and we obtain 
\beq
   t L\sim \frac{b^2\mu_1^2}{8\pi r_g^3}\left[\frac{1}{2}\left(1+\frac{\mu_1^{1/3}}{\mu_2^{1/3}}\right)\right]^4.
\eeq
A significant fraction of the kHz wave energy will be dissipated and converted to X-rays. This process can produce a detectable X-ray precursor of the merger if the stars have strong magnetic fields. For instance, a binary with $\mu_1=\mu_2=10^{32}\,$G\,cm$^3$ may generate X-ray luminosity $L_{\rm X} \sim 10^{46}\,$erg/s at $t\sim 10$\,s before the merger. Wave dissipation provides an alternative to the recently proposed precursor emission by magnetic flares \citep{Most20,Beloborodov21a}. The latter mechanism was shown to operate in binaries with anti-aligned $\boldsymbol{\mu}_1$ and $\boldsymbol{\mu}_2$; it is driven by the over-twisting of magnetic flux tubes connecting the two stars.

\subsection{Shocks from neutron star collapse}

A massive neutron star can be born (e.g. in a merger) with fast rotation, which temporarily supports it against collapse. Such objects are likely strongly magnetized and gradually lose rotation by emitting angular momentum in a magnetized wind.  The spindown can eventually lead to the collapse of the massive neutron star into a black hole, producing an electromagnetic transient \citep{Lehner12,Falcke14,Most18}. Numerical simulations of the collapse show that it launches a strong outgoing electromagnetic pulse propagating through the outer dipole magnetosphere. A monster shock will form where the pulse amplitude becomes comparable to the dipole field of the pre-collapse magnetosphere and $B^2-E^2$ reaches zero. This condition was indeed seen in vacuum and FFE simulations of the magnetosphere of a collapsing star (see Figure~10 in \cite{Lehner12}). Demonstrating shock formation and tracking its propagation requires a full MHD calculation, as shown in the present paper. The shock will dissipate a significant fraction of the outgoing electromagnetic pulse and produce a bright X-ray transient.

\medskip

I am grateful to Alex Chen for sharing their particle-in-cell simulations of strong magnetosonic waves prior to publication. This work was prompted in a large part by their results, which demonstrate shock formation through first-principles, fully kinetic calculations. I am also grateful to the anonymous referee for many useful comments, to Yuri Levin and Ashley Bransgrove for discussions of fast magnetosonic waves from quakes, and to Jens Mahlmann, Emanule Sobacchi, and Sasha Philippov for reading the manuscript and providing comments. This work is supported by NSF AST 2009453, NASA 21-ATP21-0056 and Simons Foundation \#446228.

\newpage 

\appendix

\vspace*{-0.5cm}
\section{Shock formation in a cold magnetosonic wave}
\label{cold_wave}

Below we describe analytically the flow of $C^+$ characteristics. We here focus on the part of $C^+$ flow with $E<0$ (or, equivalently, $\c<1$), which experiences strong deformation, leading to caustic formation.

\subsection{Evolution of the plasma compression factor $\c$ along $C^+$}
\label{kappa_t}

The relation between $\c$ and $r$ along $C^+$ is found from the ratio  \Eqs~(\ref{eq:beta+}) and (\ref{eq:evol}),
\beq
  \frac{dr}{d\c}=\frac{r(4\sigbg\c^3-1)}{4\sigbg\c^2(\c^2-1)} 
  \quad \Rightarrow \quad \frac{d\sigbg}{d\c}=\frac{3\sigbg\c^3-3/4}{\c^2(1-\c^2)}
  \qquad ({\rm along~}C^+),
\eeq
where we used $d\ln\sigbg/d\ln r=-3$. This is a linear differential equation for $\sigbg(\c)$. Its integrating factor is $(1-\c^2)^{3/2}$, and its solution is
\begin{eqnarray}
   \sigbg(\c) &=& \frac{3}{8(1-\c^2)^{3/2}} \left[ \frac{8}{3}\sigK + \frac{2}{\c}\sqrt{1-\c^2} \right.
    + \left. 4\arctan \left(\frac{1-\sqrt{1-\c^2}}{\c}\right) -\pi \right], 
\label{eq:sigbg_c}
\end{eqnarray}
where $\sigK$ is a constant (defined for each characteristic). Note that 
\beq
   \sigK = \lim _{\c \to 1} \left[(1-\c^2)^{3/2} \sigbg \right]
\eeq
is set at small radii  where $\sigbg\gg\sigK$. In this inner zone $|\c-1|\ll 1$ corresponds to slow plasma motions in the wave, $|\beta|\ll 1$, which is equivalent to $\Bbg\gg|E|$. Comparing the behavior of $\sigbg=\sigK(1-\c^2)^{-3/2}$ with $\sigbg=D/r^3$ (\Eq~\ref{eq:sigbg}) one can see that
\beq
\label{eq:rK}
  1-\c^2=\frac{r^2}{\rK^2}, 
  \qquad   \rK=\left(\frac{D}{\sigK}\right)^{1/3}.
\eeq 
For each $C^+$, the regime $|\c-1|\ll 1$ holds at radii $r\ll\rK$. In this zone, the radial position of $C^+$ follows vacuum propagation: $r=\rvac=c(t-\xiin)$. Note that $\rK(\xiin)$ is different for different characteristics. 

Relation~(\ref{eq:rK}) is the same as \Eq~(\ref{eq:cC+}), and so the constants $\rK$ and $\K$ are related by
\beq
   \rK(\xiin) = \frac{1}{\sqrt{-2\K(\xiin)}}.
\eeq
\Eq~(\ref{eq:rK}) holds as long as $1-\beta_+\ll 1$, which corresponds to $\c\gg\sigK^{-1/3}$. We are interested in a more general regime, where $\beta_+$ can significantly drop or even change sign, which means that the $C^+$ characteristic turns back to the star. This happens if $\c$ decreases below $(4\sigK)^{-1/3}$. In this extreme case, the variation of $r$ with $\c$ becomes non-monotonic; it occurs on a small scale $\delta r \ll r$ and can be seen only when retaining the small terms in \Eq~(\ref{eq:sigbg_c}).

Let us now find the solution for $\c(t)$ along $C^+$. Using \Eq~(\ref{eq:sigbg_c}) in \Eq~(\ref{eq:evol}), we obtain
\beq
 \label{eq:dc_dt}
  \left.\frac{d\c}{dt}\right|_{C^+} = -\frac{c\, \sqrt{1-\c^2}}{\c\,\rK}
  \left[1+\frac{(1-\c^2)^{3/2}}{4\sigK\c^3}\right]^{-1}
  \left[1+{\cal O}\left(\frac{1}{\c\sigK}\right)\right].
\eeq
Note that $\c$ monotonically decreases with $t$. This equation can be integrated: 
\beq
\label{eq:c}
   \sqrt{1-\c^2}+\frac{(1-\c^2)^2}{4\sigK \c}=\frac{c(t-\xiin)}{\rK}=\frac{\rvac}{\rK}.
\eeq
The second term on the l.h.s. makes a negligible correction to $\sqrt{1-\c^2}$ if $4\sigK\c^3\gg 1$, which corresponds to $1-\beta_+\ll 1$. Then, $\c=\sqrt{1-\rvac^2/\rK^2}=\sqrt{1+2K\rvac^2}$. If the condition $4\sigK\c^3\gg 1$ is not satisfied, one must keep both terms on the l.h.s.  This may happen only at $\c\ll 1$, and hence the deviation of $\c(t)$ from the solution $\c=\sqrt{1-\rvac^2/\rK^2}$ can be found using expansion in $\c\ll 1$. This expansion simplifies \Eq~(\ref{eq:c}) to a depressed cubic equation,
\beq
\label{eq:cubic}
     \c^3+2\left(\frac{\rvac}{\rK}-1\right)\c-\frac{1}{2\sigK}=0 \qquad \left(\frac{1}{\sigK}\ll\c\ll 1\right).
\eeq
It has one real root if $\sigK^{2/3}(1-\rvac/\rK)< 3/2^{7/3}$. Otherwise it has three real roots, but only one of them is relevant --- the root branch that gives $\c>0$ and approaches $\c=\sqrt{1-\rvac^2/\rK^2}$ when $\sigK^{2/3}(1-\rvac/\rK)\gg 1$.

The evolution of $\c$ along $C^+$ with $K<0$ may now be summarized as follows
 \begin{eqnarray}
  \c(t)=
 \left\{\begin{array}{cc}
  \displaystyle{\sqrt{1-\frac{\rvac^2}{\rK^2}} } & y \gg 1 \\
   \displaystyle{{2\sqrt{y}}\,{\sigK^{-1/3}} }\,
 \cos \left[ \frac{1}{3}\arccos\left(\frac{1}{4y^{3/2}}\right)\right] & \quad
 \displaystyle{ 2^{-4/3} <y\ll \sigK^{2/3} }  \\
 \displaystyle{{2\sqrt{y}}\,{\sigK^{-1/3}} }\,
 \cosh \left[ \frac{1}{3}{\rm arccosh}\left(\frac{1}{4y^{3/2}}\right)\right] & \quad  
  \displaystyle{0<y<2^{-4/3}} \\
    -\displaystyle{{2\sqrt{-y}}\,{\sigK^{-1/3}} }\,
   \sinh \left[ \frac{1}{3}{\rm arcsinh}\left(-\frac{1}{4|y|^{3/2}}\right)\right] & \quad  y<0   
                 \end{array}\right\},
               \qquad 
               \begin{array}{l}
               \displaystyle{ y \equiv \frac{2}{3}\,\sigK^{2/3}\left(1-\frac{\rvac}{\rK}\right) }
                   \\
               \rvac \equiv c(t-\xiin)
               \end{array}
\label{eq:kappa_C+}
\end{eqnarray}
These real expressions fully describe $\c(t)$ along $C^+$ (note that $\sigK\gg 1$ implies a large overlap of $y\gg 1$ and $y\ll \sigK^{2/3}$).

\subsection{Crossing of characteristics}
\label{crossing}

The crossing of $C^+$ characteristics is determined by the solution for their shapes $\xi_+(t)$. It is easier to solve for $\xi_+(\c)$ first, and then use the (monotonic) relation between $\c$ and $t$ along $C^+$ (\Sect~\ref{kappa_t}). The solution for $\xi_+(\c)$ can be found from the ratio of \Eqs~(\ref{eq:C+}) and (\ref{eq:evol}),
\beq
   \frac{d\xi_+}{d\c}=\frac{r}{2c\, \sigbg\c^2(\c^2-1)}.
\eeq
We substitute here $\sigbg(\c)$ (\Eq~\ref{eq:sigbg_c}) and $r=(\sigK/\sigbg)^{1/3}\rK$ to obtain a closed differential equation for $\xi_+(\c)$. Its integration (with the initial condition $\xi_+=\xiin$ at $\c=1$) yields
\beq
\label{eq:xiC+}
   \xi_+=\xiin+\frac{\rK}{2c\,\sigK\c}\left[(1-\c)^2 + {\cal O}\left(\frac{1}{\sigK\c}\right)\right].
\eeq
One can think of the initial value $\xiin$ as a Lagrangian coordinate in the flow of $C^+$ characteristics and quantify deformation of the wave profile using ``strain'' $\partial\xi_+/\partial\xiin$ evaluated at $t=const$. A reduction of $\partial\xi_+/\partial\xiin$ below unity means compression of the profile, i.e. steepening of the wave. The caustic appears where $\partial\xi_+/\partial\xiin$ vanishes. This will occur in the region of $\sigK^{-1}\ll\c\ll 1$, and in this region the expression for $\partial\xi_+/\partial\xiin$ simplifies to
\beq
   \left.\frac{\partial\xi_+}{\partial\xiin}\right|_t \approx 1 -\frac{2 (1+\drK)}{4\sigK\c^3+1},
    \quad {\rm where} \quad \drK\equiv \frac{d\rK}{d\xiin} =\rK^3 \dot{\K}, \qquad \dot{K}\equiv\frac{dK}{d\xiin}.
\eeq
Here, we used \Eq~(\ref{eq:cubic}) to express $(\partial\c/\partial\xi_+)_t$ and retained only the leading terms of the expansion in $\c\ll 1$. Thus, for each characteristic $C^+$ with $\K<0$ we find that $\partial\xi_+/\partial\xiin$ vanishes when $\c=\cv$, where $\cv$ satisfies the relation
\beq
\label{eq:cv}
   \cv^3(\xiin)\approx \frac{2\dot{r}_\star+1}{4\sigK} \qquad (\cv\ll 1).
\eeq  
The condition $\c=\cv$ corresponds to time $\tv$ determined by \Eq~(\ref{eq:cubic}):
\beq
  \frac{t}{\rK} = 1 - \frac{\c^2}{2} + {\cal O}(\c^4) +\frac{1+{\cal O}(\c)}{4\sigK\c} 
  \quad \Rightarrow \quad 
   \tv(\xiin)\approx \frac{\rK}{c}\left(1-\frac{\cv^2}{2}+\frac{1}{4\sigK\cv}\right).
\label{eq:tv}
\eeq
The small difference $\tv-\rK/c$ is determined by the small $\cv\ll 1$, which is given in the leading order by \Eq~(\ref{eq:cv}).

The caustic forms on the characteristic $C^+$ for which $\partial\xi_+/\partial\xiin=0$ is reached first, i.e. where $\tv(\xiin)$ is minimum. This characteristic (labeled as $\xiinc$) satisfies the condition $dt_{\rm v}/d\xiin=0$. Using \Eq~(\ref{eq:cv}) we find
\beq
   \frac{d\cv}{d\xiin}=\frac{\ddot{r}_\star}{6\sigK\cv^2}+\frac{\cv\dot{r}_\star}{\rK},
  \quad {\rm where} \quad \ddot{r}_\star=\frac{3\dot{r}_\star^2}{\rK}-\frac{\rK \ddot{\K}}{2\K},
\eeq 
and then obtain the condition $d\tv/d\xiin=0$ in the form,
\beq
   \frac{d\tv}{d\xiin}=(1+\drK)\left(1+\frac{\rK^2\ddot{\K}}{24\sigK^2\cv^2 \K}\right) + Z = 0.
\eeq
The terms collected in $Z(\cv)$ are small, and we find a simple expression for the fluid compression factor at the caustic, $\cc\equiv\cv(\xiinc)$ (one can show that $Z(\cc)\ll 1+\drK$):
\beq
\label{eq:cc1}
   \cc^4\approx -\frac{\rK^2\ddot\K}{24\sigK^2\K}.
\eeq

For example, consider waves with a harmonic profile $\K(\xiin)$. Let us define $\psi\equiv\omega\xiin$ so that $\K(\psi)=\Km\sin\psi$. Then,
\beq
   \rK=\frac{\Rm}{\sqrt{-\sin\psi}}, \qquad \sigK=\sigm(-\sin\psi)^{3/2}, \qquad 
   \drK=\frac{\Rm \omega\cos\psi}{2c(-\sin\psi)^{3/2}},
   \qquad
   \frac{\ddot\K}{\K}=-\omega^2.
\eeq
The caustic develops at a Lagrangian coordinate $\psi_c=\omega\xiinc$,
and \Eqs~(\ref{eq:cv}), (\ref{eq:cc1}) give
\beq
   4\, (-\sin\psic)^{3/2}\cc^3=\frac{\zeta\cos\psic}{(-\sin\psic)^{3/2}}+\frac{1}{\sigm}, \qquad
  \cc^4 = \frac{\zeta^2}{24\sin^4\!\psic}, \quad {\rm where} \quad \zeta \equiv\frac{\omega\Rm}{c\sigm}.
  \label{eq:cc3}
\eeq
These equations hold if $\cc\ll 1$, which requires $\zeta\ll 1$.
Excluding $\cc$ from \Eqs~(\ref{eq:cc3}) we obtain 
\beq
\label{eq:psic}
     \cos\psic=\frac{4\sqrt{\zeta}}{24^{3/4}} - \frac{c}{\omega\Rm}\,(-\sin\psic)^{3/2}.
\eeq
We consider only short waves, $\omega\Rm/c\gg 1$. \Eq~(\ref{eq:psic}) then implies $|\cos\psic|\ll 1$, and so in the leading order one can set $\sin\psic\approx -1$ in \Eqs~(\ref{eq:cc3}) and (\ref{eq:psic}). This gives the relations stated in \Eq~(\ref{eq:caustic}).

The caustic forms at time $\tc=\tv(\xiinc)$ given by \Eq~(\ref{eq:tv}). Expanding $\rK/\Rm=(1-\cos^2\psi_c)^{-1/4}\approx 1+(1/4)\cos^2\psi_c$ and using \Eq~(\ref{eq:caustic}) we obtain \Eq~(\ref{eq:tc}).

The caustic position in the wave, $\xi_c$, may be found using the relation between $\xi_+$ and $\c$ along $C^+$ (\Eq~\ref{eq:xiC+}). It gives $\xi_c\approx \xiin^c + \rK/2c\sigK\cc$, which leads to \Eq~(\ref{eq:xic}).


\section{Shock micro-structure}
\label{shock_structure}

The flow is composed of two streams, $e^+$ and $e^-$, which have different (symmetric) trajectories. The $e^\pm$ flow is everywhere neutral, i.e. its net charge density is zero. As the flow enters the shock, it develops a transverse electric current $\bj$, created by the opposite $e^\pm$ motions along $\bE$. This current controls the self-consistent change of the magnetic field across the shock. In this section $t$, $\bE$, $\bB$, $\bb$, $u^\alpha$, and $n$ will denote quantities measured in the shock rest frame (elsewhere in the paper this notation is used for the lab-frame quantities). 

We will use local Cartesian coordinates $x,y,z$ with the $x$-axis along $\bE$ and the $z$-axis along the upstream flow. Then, the magnetic field $\bB$ is along $y$. The upstream quantities ahead of the shock will be denoted with subscript ``u''. The upstream flow will be approximated as cold, i.e. the $e^\pm$ are at rest in the plasma drift frame (the  frame $\tilde{\KF}$ where the electric field vanishes, $\tilde{E}=0$). In this section, we use the units of $c=1$.

The steady flow is described by quantities that depend only on $z$. In this approximation, all time derivatives vanish, $\partial_t=0$, and hence $\nabla\times\bE=-\partial_t\bB=0$. In addition, $\nabla\cdot\bE=0$ as the net charge density is zero. This implies  
 \beq
   \bE(z)=const=\bE_{\rm u}=(\Eu,0,0).
\eeq
The $e^\pm$ streams ahead of the shock have equal velocities 
\beq
   \bb_{\rm u}^\pm=(0,0,\betau), \qquad \betau=\frac{\Eu}{\Bu}.
\eeq
The straight inertial motion of $e^\pm$ describes the cold upstream flow with no gyration in the drift frame $\tilde{\KF}$; the $e^\pm$ experience zero force $\pm e(\bE_{\rm u}+\bb_{\rm u}^\pm\times\bB_{\rm u})=0$.

\subsection{Shock structure equations}

As the plasma crosses the shock, the $e^\pm$ four-velocities $u^\alpha_\pm=(\gamma_\pm, \boldsymbol{u}_\pm)$ and $B$ change from their upstream values. The profiles $u_\pm^\alpha(z)$ and $B(z)$ obey the equations:
\beq
\label{eq:dynamic}
    m\, \beta_z \frac{d\gamma_\pm}{dz} = \pm e\bE\cdot\bb_\pm -\dot{\E}_e, 
   \qquad
    m\, \beta_ z \frac{d\boldsymbol{u}_\pm}{dz} = \pm e\left(\bE+\bb_\pm\times\bB\right)-\boldsymbol{f}_\pm, 
    \qquad
  \nabla\times\bB = 4\pi \bj,
\eeq
where $\dot{\E}_e $ is the power radiated by the ultra-relativistic particle and $\boldsymbol{f}_\pm=\bb_\pm\dot{\E}_e$ is the radiated momentum. Similar equations, but without radiation reaction, were previously used to describe magnetosonic solitons, which have no dissipation \citep{Kennel76,Alsop88}. 

The particle entering the shock radiates because it experiences acceleration $(\pm e/\me)(\bE+\bb_\pm\times\bB)$. The power radiated by a particle with four-velocity $(\gamma,\boldsymbol{u}$) is given by (e.g. \citealt{Landau75}),
\beq
 \dot{\E}_e = \frac{\sT}{4\pi}\left[(\gamma\bE+\boldsymbol{u}\times\bB)^2-(\boldsymbol{u}\cdot\bE)^2\right] 
  = \frac{\sT}{4\pi}\left[(\gamma B- u_zE)^2+E^2-B^2\right],
\eeq
where $\sT=8\pi r_e^2/3$ is the Thomson cross section.

\Eqs~(\ref{eq:dynamic}) have a symmetry: the $e^\pm$ develop $\beta_x^-=-\beta_x^+$, $\beta_z^-=\beta_z^+$, and keep $\beta_y^-=\beta_y^+=0$. Therefore, it is sufficient to solve for the four-velocity of $e^+$. Below it will be denoted by $u^\alpha$, omitting subscript ``$+$''. The symmetry also implies that the electric current has the form $\bj=(j,0,0)$, where $j=en\beta_x$, $\bb\equiv\bb_+$ and $n=n_++n_-$  is the local plasma density. It is related to the uniform particle flux $F=n\beta_z=n_{\rm u} \betau$,
\beq
   j=eF\,\frac{\beta_x}{\beta_z} = e n_{\rm u} \betau \,\frac{\beta_x}{\beta_z}.
\eeq
The current $j$ is expected to be negative, since $B$ is compressed in the shock: $dB/dz=-4\pi j>0$. This requires $\beta_x<0$. The solution presented below indeed shows that the $e^+$ flow entering the shock develops negative acceleration along $x$: $\beta_z du_x/dz=(e/m)(E-\beta_zB)<0$. This is consistent with energy conservation: the compression of $B$ occurs at the expense of the flow kinetic energy, and hence $\bE\cdot\bj<0$ (energy is extracted from the plasma). 

Note that $\gamma$ and $u_z$ determine all components of four-velocity $u^\alpha$  ($u_x$ is found from $u^\alpha u_\alpha=-1$ using $u_y=0$), and hence also determine $\beta_x$ and $\beta_z$. As a result, we have three coupled equations for three unknowns $\gamma$, $u_z$, $B$. The flow solution has three parameters: $n_{\rm u}$, $\Bu$, $\betau$. Given these parameters, the system of equations~(\ref{eq:dynamic}) can be integrated numerically.  As shown below, the solution also admits a simple analytical approximation.

A normal collisionless shock with no radiation reaction ($\chi=0$) would be described by two dimensionless parameters: $\gu=(1-\betau^2)^{-1/2}$ and $\sigu=\Bu^2/4\pi m n_{\rm u}\gu$. Its downstream moves with Lorentz factor $\gd\approx\sqrt{\sigu}$ relative to the shock. The shock structure is normally presented using fields normalized to $\Bu$ and distance normalized to the Larmor scale $\rL\equiv m\gu/e\Bu$. By contrast, the presence of radiation reaction in \Eqs~(\ref{eq:dynamic}) implies that $\Bu$ cannot be removed by the normalization of fields, since $\dot{\E}_e$ is quadratic in the fields. An additional scale $r_e=e^2/m$ appears in the problem, and the value of $\Bu$ controls the ratio $r_e/\rL$. To disentangle this problem, it helps to introduce new variables:
\beq
  g\equiv \frac{\gamma}{\gu}, 
   \quad
    w\equiv \frac{\sigu(\gamma-u_z)}{\gu}, \quad b\equiv\sigu\left(\frac{B}{\Bu}-1\right).
\eeq
Instead of coordinate $z$ we will use dimensionless $\tt$ defined by 
\beq
  d\tt=\frac{e\Bu}{m\gu \sqrt{\sigu}}\,\frac{dz}{\beta_z}.
\eeq
\Eqs~(\ref{eq:dynamic}) then become
\beq
\label{eq:dynamic1}
  \frac{dg}{d\tt} = -\betau \FF - \chi\R, \qquad
   \frac{d\ww}{d\tt} = -\FF\left(b+\frac{\eps}{2}\right) - \ww \chi \R, \qquad
    \frac{db}{d\tt} = \betau \FF,
\eeq
where
\beq
\label{eq:chi}
    \eps\equiv \frac{\sigu}{\gu^2}, \qquad 
    \chi\equiv\frac{\sT\Bu\gu^2}{4\pi e\,\sigu^{3/2}}=\frac{2}{3\eps^{3/2}} \frac{r_e}{\rL},
\eeq
\begin{eqnarray}
\label{eq:F_}
    \FF(g,\ww,b)&\equiv& \frac{1}{g}\sqrt{\ww\left(2g-\frac{\ww}{\sigu}\right)-\eps}
    = -\beta_x\sqrt{\sigu}, \\
    \R(g,\ww,b) &\equiv& \left(\ww+\frac{g-\ww/\sigu}{1+\betau}\,\eps\right)^2-\eps^2+2bg\ww + \frac{2bg(g-\ww/\sigu)}{1+\betau} \eps-2b\eps+g^2b^2.
\label{eq:R}
\end{eqnarray} 

Note that the presented form of the flow equations (\ref{eq:dynamic1}) uses $\betau,\eps, \chi$  as the three parameters of the shock instead of $\gu,\sigu,\Bu$. The parameter $\sigu$ appears only in the functions $\FF$ and $\R$, where it enters through the term $g-\ww/\sigu$. We will verify below that $\ww/\sigu\ll g$ throughout the shock transition, and so $\sigu$ drops out. 

The choice of $\betau,\eps, \chi$ as the independent parameters significantly simplifies the problem. For strong ultra-relativistic shocks of interest, (i) one can set $\betau=1$ and (ii) $\eps$ turns out so small that it effectively drops out. As a result, $\chi$ remains as the only important parameter (while $\sigu$ and $\gu$ enter through the definition of variables $g, \ww, b$). 

In the limit of $\chi\rightarrow 0$, the autonomous system~(\ref{eq:dynamic1}) admits an immediate analytical solution. Its first integrals $g(b)$ and $\ww(b)$ are
\beq
\label{eq:no_loss}
   g=1-b, \qquad \ww=\frac{b(b+\eps)+\eps\betau}{2\betau} \qquad (\chi=0),
\eeq
where we took into account that the solution starts in the upstream with $b=0$, $g=1$, $\ww=\eps/2$. The first integrals correspond to the fluxes of specific energy and momentum staying uniform along $z$, as losses are neglected. 

The monster shocks discussed in this paper propagate in the qualitatively different regime of $\chi\gg 1$. The value of $\chi$ is controlled by the shock strength developing in the kHz wave; it will be evaluated in \Sect~\ref{analytical}.

\subsection{Boundary conditions}
\label{boundary}

In order to integrate \Eqs~(\ref{eq:dynamic1}), one needs boundary conditions $g=g_0$, $\ww=\ww_0$, $b=b_0$ at some $\tt_0$. We will set the boundary conditions in the upstream, ahead of the shock. Far into the upstream the flow is unaware of the shock and has $g=1$, $\ww= \eps/2$, $b=0$, which corresponds to $\gamma=\gu$, $u_z=\gu\betau$, $B=\Bu$. However, one cannot impose these simple boundary conditions, because \Eqs~(\ref{eq:dynamic1}) have a trivial uniform solution with  these values. The shock solution must have an infinitesimal deviation from the uniform solution at $\tt_0\rightarrow -\infty$. It is obtained by choosing $g_0,\ww_0,b_0$ slightly perturbed from $1, \eps/2, 0$. 

The upstream flow with $g,\ww,b=1,\eps/2,0$ has $\R=0$, and the deviation of $g,\ww,b$ from $1,\eps/2,0$ begins with negligible losses ($\chi\R\ll \FF$). Hence, one can choose $\tt_0$  where the flow still satisfies the adiabatic solution~(\ref{eq:no_loss}). Only one of the three perturbed quantities at $\tt_0$ is independent: by choosing a small perturbation $b_0\neq 0$, we also determine $g_0$ and $\ww_0$ according to \Eq~(\ref{eq:no_loss}). 
Furthermore, changing $b_0$ is equivalent to shifting $\tt_0$ in the profile of $b(\tt)$. This shift is irrelevant, since the problem has translational symmetry along $\tt$. Thus, no new free parameters are introduced by the choice of boundary conditions. The shock transition will follow a unique solution apart from the arbitrary shift along the $\tt$-axis.

In practice, $b_0$ does not need to be infinitesimal. It is sufficient to choose any finite small $b_0$ in the zone where the adiabatic solution~(\ref{eq:no_loss}) still holds, i.e. where $\dot{\E}_e\ll e|\bE\cdot\bb|$. This requires 
\beq 
\label{eq:initial}
  \chi \R(g_0,\ww_0,b_0) \ll \FF(g_0,\ww_0,b_0).
\eeq
For shocks of main interest ($\betau\rightarrow 1$ and small $\eps$, $\sigu^{-1}$) we find, using \Eq~(\ref{eq:no_loss}):
\beq
   \R\approx b^2\left(1-\frac{b}{2}\right)^2, 
  \;\; \FF\approx \frac{b}{\sqrt{1-b}} \quad (\dot{\E}_e\ll e|\bE\cdot\bb|)
\eeq
The adiabatic condition~(\ref{eq:initial}) is satisfied if $b_0\ll\chi^{-1}$. In the numerical models presented below we set $b_0=\eps^{1/2}$. It satisfies condition~(\ref{eq:initial}) for shocks with $\eps\ll\chi^{-2}$.

One can also see here that shocks with $\chi\ll 1$ have negligible losses ($\chi\R\ll \FF$) throughout the shock transition, as $g$ drops from unity and $b=1-g$ changes from 0 to $b\sim 1$. By contrast, $\chi\gg 1$ implies almost immediate onset of radiation reaction as the flow enters the shock.

\subsection{Numerical solution}
 \label{numerical_structure}

Numerical solutions for shocks with different values of $\chi$ are shown in Figure~8. They are obtained in the limit of $\sigu\gg 1$ and $\eps\ll 1$ (we used $\sigu=10^3$ and $\eps=10^{-16}$ when integrating \Eqs~(\ref{eq:dynamic1}), however the result is independent of these choices). Note that the flow everywhere keeps $\beta_z\approx 1$, and hence $\tt\propto z$. One can see that in the shock the normalized Lorentz factor of the flow $g(\tt)=\gamma/\gu$ decreases while the magnetic field becomes compressed: $b(\tt)$ grows from zero up to the final downstream value. The relation between $g$ and $b$ across the shock transition is shown in Figure~8.

\begin{figure}[t]
\begin{center}
\includegraphics[width=0.47\textwidth]{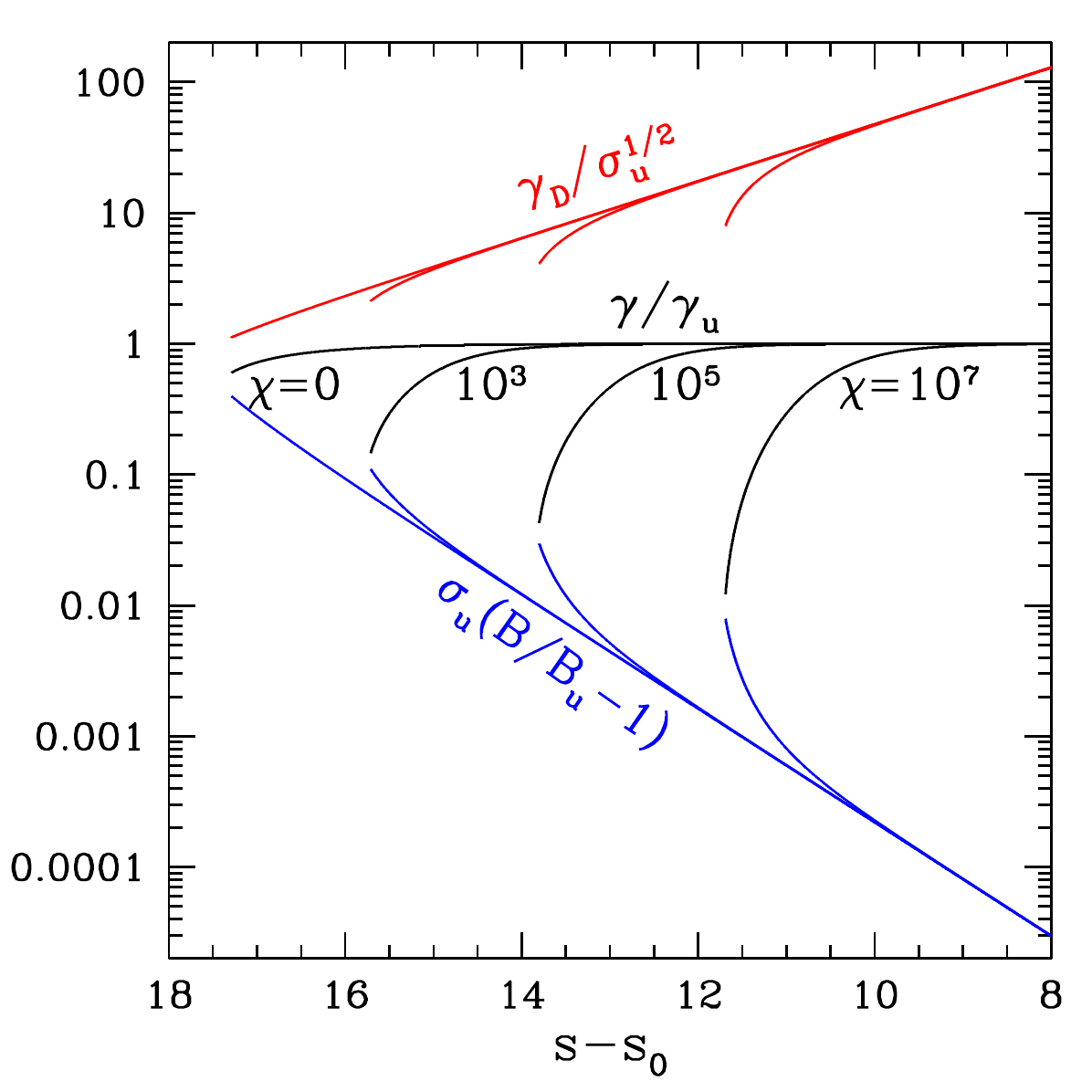}
\includegraphics[width=0.47\textwidth]{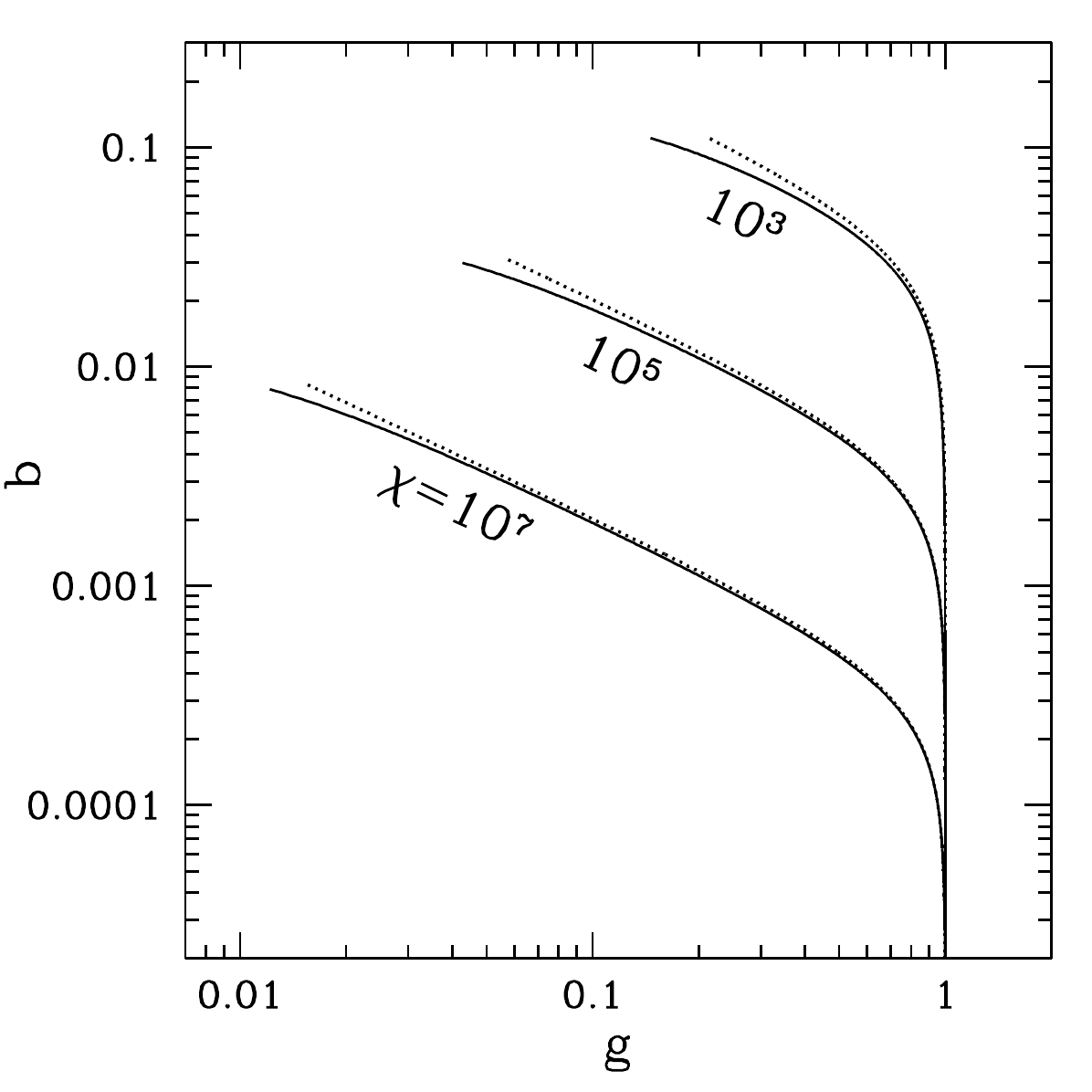}
\caption{Shock structure solution in the approximation of steady $e^\pm$ streams (viewed in the shock rest frame). The solution was calculated for $\chi=0,10^3,10^5,10^7$. {\it Left:} For each case, we show $g(\tt)$ (black), $b(\tt)$ (blue), and $\gamma_{\rm D}(\tt)=B/\sqrt{B^2-E^2}$ normalized to $\sqrt{\sigu}$ (red). Flows with $\chi\gg 1$ radiate most of their energy before developing gyration; this results in $\gamma\ll\gu$. Each solution ends where the downstream begins, i.e. where the $e^\pm$ streams develop gyration (then they become unstable, thermalizing their gyration energy in the drift frame). The drift Lorentz factor $\gamma_{\rm D}$ at the end point represents the downstream speed relative to the shock.
{\it Right:} Relation between $g$ and $b$. The approximate analytical solution (\Eq~\ref{eq:b_g}) is shown by the dotted curve and compared with accurate numerical integration (solid curve). Both numerical and analytical solutions are plotted until the $e^\pm$ streams reach $|\tilde{\beta}_x|=1/2$, which corresponds to $60^{\circ}$ gyration in the drift frame.}
\end{center}
\label{fig:structure}
 \end{figure}

The presented solutions end where the flow develops gyration in the drift frame $\tilde{\KF}$. Beyond this point, the maser instability is expected to destroy the coherent gyration of the $e^\pm$ streams \citep{Langdon88}. This instability is not followed by our calculation, and this is not needed to see the effect of radiation reaction on the shock structure. After the development of gyration radiative losses become normal synchrotron emission, which carries away zero momentum in the drift frame and no longer affects the bulk motion of the plasma.

We set the end point $\tt_{\rm end}$ of the coherent stream solution where the $e^\pm$ velocity vector in the drift frame, $\tilde{\bb}_\pm$, has rotated by $60^{\circ}$ ($\tilde{\beta}_x^{\pm}=\mp 1/2$). We define the approximate downstream quantities at this point. In particular, the drift Lorentz factor $\gD=B/\sqrt{B^2-E^2}$ at $\tt_{\rm end}$ approximately represents the Lorentz factor of the downstream relative to the shock, $\gd$. Figure~\ref{fig:jump} shows the measured $\gd$ as a function $\chi$.\footnote{Different choices for the end point definition (up to $\tilde{\beta}_x^{\pm}=\mp \tilde{\beta}\approx 1$) affect the measured $\gd$ by tens of percent, a modest correction compared with the main effect --- radiation reaction before the development of gyration. The exact $\gd$ may be found with the full kinetic plasma simulation of the shock.} 
For shocks with $\chi\gg 1$, the numerical result approximately follows the relation $\gd/\sqrt{\sigu}\approx \chi^{1/7}$. It can also be derived analytically, as shown below.

\subsection{Approximate analytical solution}
\label{analytical_structure}

The radiation-reaction transition admits a simple analytical description. A key fact simplifying the analytical approximation is that the term $\ww\chi\R$ in $dw/ds$  (\Eq~\ref{eq:dynamic1}) remains small throughout the shock transition. When this term is neglected, the equations for $dw/ds$ and $db/ds$ can be integrated for $\ww(b)$. The resulting relation $\ww(b)$ is the same as in the adiabatic solution~(\ref{eq:no_loss}). At $b\gg \eps$ it gives
\beq
\label{eq:w1}
   \ww\approx \frac{b^2}{2}.
\eeq
Note also that the functions $\R$ and $\FF$ entering the dynamical equations can be simplified. Neglecting the small terms $\propto\eps$ in \Eq~(\ref{eq:R}), we obtain
\beq
\label{eq:R2}
   \R\approx \left(\ww+gb\right)^2 \approx b^2 \left(\frac{b}{2}+g\right)^2, 
   \qquad \FF\approx \frac{b}{\sqrt{g}},
\eeq
where we used $\ww/\sigu\ll g$ to further simplify $\FF$. Like \Eq~(\ref{eq:w1}), \Eqs~(\ref{eq:R2}) approximately hold at $b\gg \eps$, practically throughout the entire shock transition.

With these preliminary remarks, we can describe the shock structure as follows. The shock has two zones:

{\bf Adiabatic zone ($g_1<g<1$)}.
Here, the upstream flow enters the shock by following the solution~(\ref{eq:no_loss}) for $g(b)$ and $\ww(b)$. The adiabatic zone ends when the radiative term becomes important in $dg/ds$ (\Eq~\ref{eq:dynamic1}), i.e. when $\chi\R$ becomes comparable to $\FF$. This occurs at $b_1\sim \chi^{-1}\ll 1$. By this moment, the flow has lost only a small fraction $1-g_1$ of its initial kinetic energy, which went into the slight compression of the magnetic field,
\beq
\label{eq:g1}
   b_1\approx 1-g_1 \approx \frac{1}{\chi}\ll 1.
\eeq

{\bf Radiation-reaction zone ($g_2\ll g \ll g_1$)}.
Here, $\chi\R\gg \FF$ and the flow deceleration is controlled by radiative losses. The solution for $b(g)$ can be found from the ratio of equations $dg/d\tt\approx -\chi\R$  and $db/d\tt=w$ (\Eq~\ref{eq:dynamic1}). Using $\R\approx g^2b^2$ (this holds because $gb\gg \ww$, as verified below), we find 
\beq
\label{eq:b_g_}
   \frac{3}{4}\chi (b^2-b_1^2)  \approx g^{-3/2}-g_1^{-3/2}.
\eeq 
We have chosen the integration constant so that $b(g_1)$ matches $b_1$ (\Eq~\ref{eq:g1}). However, this boundary condition becomes unimportant in the radiation-reaction zone, where $b\gg b_1$ and $g^{-3/2}\gg 1-g_1$. So, the solution simplifies to
\beq
\label{eq:b_g}
   \frac{3}{4}\chi b^2 \approx g^{-3/2}-1.
\eeq
This result is in excellent agreement with the accurate numerical solution (Figure~8).

The key approximation of this derivation was the neglect of $\ww\chi\R$ compared with $b\FF$ in $dw/ds$ (\Eq~\ref{eq:dynamic1}). Evaluating the ratio
\beq
   \frac{\ww\chi\R}{b\FF}\approx \frac{2}{3}\,g\left(1-g^{3/2}\right),
\eeq
one can see that it has the maximum of $(2/5)^{5/3}\approx 0.2$, a small value. One can also verify that $\ww/\sigu\ll g$ and $gb/\ww\approx 2g/b\approx 3\chi\R/\FF\gg 1$; this vindicates the approximations $\FF\approx b/\sqrt{g}$ and $\R\approx g^2b^2$.

The radiation-reaction zone ends where the term $\chi\R$ stops being dominant in $dg/ds$. This transition may be defined e.g. by $\chi\R\approx 2\FF$, which occurs at
\beq
   g_2\approx\left(\frac{3}{\chi}\right)^{2/7} \ll 1.
\eeq
As the radiation-reaction zone ends at $g\sim g_2$, the gyrating downstream begins. The development of gyration can be seen by transforming the $e^+$ stream velocity to the drift frame. We find the drift speed $\beta_{\rm D}=E/B=\betau\Bu/B$ and then the drift Lorentz factor
\beq
  \gD=\frac{B}{\sqrt{B^2-E^2}}
   =\frac{1+b/\sigu}{\sqrt{(1+b/\sigu)^2-\betau^2}}
   \approx\sqrt{\frac{\sigu}{2b}}.
\eeq
The particle Lorentz factor measured in the drift frame is
\beq
   \tilde{\gamma}\approx \gD(\gamma-\beta_{\rm D} u_z)
   \approx \gu\sqrt{\frac{b}{2\sigu}}\left(g+\frac{b}{2}\right).
\eeq
The $e^\pm$ streams are ultra-relativistic in the drift frame, i.e. $\tilde\beta\approx 1$, and $\tilde{\beta}_x\approx \sin\tilde{\psi}$ determines their deflection angles $\pm\tilde{\psi}$ from the $z$-axis. Taking into account that $\tilde{u}_x=u_x$, we find 
\beq
   \tilde{\beta_x}=\frac{u_x}{\tilde{\gamma}}=-\frac{\gamma\FF}{\tilde{\gamma}\sqrt{\sigu}}
   \approx -\frac{\sqrt{2gb}}{g+b/2}.
\eeq
Substituting here the approximate solution~(\ref{eq:b_g}) for $b(g)$, one can see the development of gyration with decreasing $g$. In particular, one can see that $|\tilde{\beta}_x|\sim 1$ at  $g\sim g_2$, confirming the formation of a gyrating downstream.

Finally, we can evaluate the shock jump condition of main interest: the downstream Lorentz factor in the shock rest frame, $\gd$ (denoted as $\gd'=\gamma_{\rm sh|d}$ in the main text). It approximately equals $\gD(g_2)$, which gives
\beq
\label{eq:jump_}
   \frac{\gd}{\sqrt{\sigu}}\approx \chi^{1/7} \qquad (\chi\gg 1).
\eeq

\newpage

\bibliography{ms}

\end{document}